\documentclass[12pt, draftclsnofoot, onecolumn]{IEEEtran}
\IEEEoverridecommandlockouts
\usepackage{cite}
\usepackage{bbm}
\usepackage{amsmath,amssymb,amsfonts,amsthm}
\usepackage{algorithmic}
\usepackage{graphicx}
\usepackage{textcomp}
\usepackage{xcolor}
\usepackage{authblk}
\usepackage{stfloats}
\usepackage{soul}
\theoremstyle{definition}
\newtheorem{prop}{Proposition}

\usepackage{cleveref}
\usepackage{subcaption}
\usepackage{mathtools}
\usepackage{refcount}

\newcommand\myeq{\stackrel{\mathclap{\normalfont\mbox{(*)}}}{=}}

\DeclareMathOperator{\sinc}{sinc}

\def\BibTeX{{\rm B\kern-.05em{\sc i\kern-.025em b}\kern-.08em
    T\kern-.1667em\lower.7ex\hbox{E}\kern-.125emX}}
    \renewcommand{\qedsymbol}{$\blacksquare$}

\makeatletter
\newcommand\footnoteref[1]{\protected@xdef\@thefnmark{\ref{#1}}\@footnotemark}
\makeatother

\begin{document}
\newcounter{cor_cnt}
\setcounter{cor_cnt}{0}
\newtheorem{cor}{Corollary}
\bstctlcite{IEEEexample:BSTcontrol}
    %\pagenumbering{gobble}
\title{Performance Analysis of Quantized Uplink Massive MIMO-OFDM With Oversampling Under Adjacent Channel Interference}
\author{Ali Bulut \"{U}\c{c}\"{u}nc\"{u},~\IEEEmembership{Student Member,~IEEE,}
{Emil Bj\"{o}rnson,~\IEEEmembership{Senior Member,~IEEE,}}
{H\r{a}kan Johansson,~\IEEEmembership{Senior Member,~IEEE,}}
{Ali \"{O}zg\"{u}r Y{\i}lmaz,~\IEEEmembership{Member,~IEEE,}}
{and~Erik G. Larsson,~\IEEEmembership{Fellow,~IEEE}}
\thanks{This version of the paper has been accepted for publication in IEEE Transactions on Communications. The digital object identifier of the accepted paper is 10.1109/TCOMM.2019.2954512.}
\thanks{The work of A. B. \"{U}\c{c}\"{u}nc\"{u} was supported by
the ASELSAN Graduate Scholarship for Turkish Academicians. \newline\indent
A. B. \"{U}\c{c}\"{u}nc\"{u} and A. \"{O} Y{\i}lmaz are with the Department
of Electrical and Electronics Engineering, Middle East Technical University, 06800, Ankara,
Turkey, (e-mail: \{ucuncu, aoyilmaz\}@metu.edu.tr)}% <-this % stops a space
\thanks{E. Bj\"{o}rnson, H. Johansson and E. G. Larsson are with the Department of Electrical Engineering (ISY), Link\"{o}ping University, SE-581 83 Link\"{o}ping, Sweden (e-mail:\{emil.bjornson, hakan.johansson, erik.g.larsson\}@liu.se)}}% <-this % stops a space}

%\affil[1]{Dept. of Electrical and Electronics Engineering, Middle East Technical University, Ankara, Turkey}
%\affil[2]{Dept. of Electrical Engineering, Link\"{o}ping University, Link\"{o}ping, Sweden}
%\affil[ ]{E-mail: ucuncu@metu.edu.tr, \{emil.bjornson, hakan.johansson, erik.g.larsson\}@liu.se, aoyilmaz@metu.edu.tr}
%\renewcommand\Authands{ and }

\maketitle

%\title{Paper Title*\\
%\author{\IEEEauthorblockN{Ali Bulut \"{U}\c{C}\"{U}NC\"{U}, Emil Bj\"{o}rnson, H\r{a}kan Johansson, Erik G. Larsson}

\begin{abstract}
Massive multiple-input multiple-output (MIMO) systems have attracted much attention lately due to the many advantages they provide over single-antenna systems. Owing to the many antennas, low-cost implementation and low power consumption per antenna are desired. To that end, massive MIMO structures with low-resolution analog-to-digital converters (ADC) have been investigated in many studies. However, the effect of a strong interferer in the adjacent band on quantized massive MIMO systems have not been examined yet. In this study, we analyze the performance of uplink massive MIMO with low-resolution ADCs under frequency selective fading with orthogonal frequency division multiplexing in the perfect and imperfect receiver channel state information cases. We derive analytical expressions for the bit error rate and ergodic capacity. We show that the interfering band can be suppressed by increasing the number of antennas or the oversampling rate when a zero-forcing receiver is employed.\end{abstract}


\section{Introduction}
\label{sec:intro}
Massive multiple-input multiple-output (MIMO) systems provide high spectral and energy efficiency with relatively simple signal processing, promoting their utilization in modern communication systems \cite{Marzetta1088544,larsson_MM_survey,what_will_5G_be, Enrgy_sp_eff_Ngo}. However, due to the large number of antennas, low-cost hardware and low power consumption per antenna may be necessary owing to feasibility reasons \cite{björnson_critical_quest,björnson_hrdw_scaling_law}. 

With regards to the employment of low-cost and power efficient hardware, low-resolution analog-to-digital converters (ADC) can be beneficial \cite{jacobsson_aggragate,jacobsson_recent,jacobsson}. It is known that the power consumption of ADCs is increased by 2 to 4 times with each bit increase \cite{murmann_extra_dB}. Among the low-resolution ADCs, one-bit ADC has the lowest possible complexity (it is composed of a single comparator; and no automatic gain control (AGC) unit is needed) and power consumption, which is why it has been analyzed widely in the massive MIMO literature \cite{larsson_MM_1bitADC,jacobsson,Li_perf_analysis,Mollen_wideband,nML_MM_MIMO, jacobsson_recent,Studer_ML_optim,mixed_adc,heat_cap_MIMO_quant,ICC_own,TWC_own}.
In particular, \cite{larsson_MM_1bitADC,Li_perf_analysis,Mollen_wideband,jacobsson,jacobsson_recent,mixed_adc,heat_cap_MIMO_quant} analyze uplink massive MIMO systems with one-bit or low-resolution ADCs in terms of capacity, error rate and achievable rate, and propose channel estimation algorithms. Temporal oversampling for such systems is proposed in \cite{TWC_own,ICC_own,VTC_wideband_own,VTC_seq_lin_rec} for single carrier transmission, along with a performance analysis. Many other studies have proposed optimal/near optimal non-linear detectors for uplink massive MIMO systems with low-resolution ADCs (e.g. \cite{Choi_quantized_rec,nML_MM_MIMO,Studer_ML_optim}). Another study \cite{jacobsson_aggragate} has analyzed the aggregate effect of an ADC, non-linear power amplifier, and phase noise on uplink massive MIMO with orthogonal frequency division multiplexing (OFDM).

%Due to the fact that there is some similarity between the analysis of systems with ADCs and digital-to-analog converters (DAC), the studies dealing with downlink massive MIMO systems with quantized DACs will also be mentioned. For systems with 1-bit and low-resolution DACs, linear precoders are proposed, some of which are working close to infinite resolution DACs (e.g. \cite{Saxena_one_bit,Li_downlink_one_bit,Jacobsson_quant_prec}). For 1-bit DACs, non-linear precoding techniques that work better than the linear precoders are obtained in \cite{Jacobsson_quant_prec,swindlehurst2017minimum}. Recently, \cite{jacobsson_lin_prec} has performed a performance analysis of a MIMO-OFDM system with quantized DACs for frequency selective channel.

%%Moreover, \cite{jacobsson_one_bit_prec} and \cite{jacobsson_1bit_non_lin_prec} propose precoding techniques for one-bit quantized massive MIMO scenario and provide the performance analysis for the proposed schemes. 

In none of the aforementioned studies, the performance of heavily quantized massive MIMO with an interferer in an adjacent channel is examined. However, such interference can be at significant levels due to near/far effect in a communication system in which users in the adjacent frequency band may be much closer to the receiver than the users in the desired band, thus, their signal may not be adequately suppressed by the receivers intending to extract the signals in the desired band. In fact, having the dynamic range to mitigate such interference is a key reason for using high-resolution ADCs in current systems \cite{berger_dynamic_range}. Since distortion is large with low-resolution ADCs, there is a risk that such systems are practically nonoperational.

This study is the first to analyze heavily quantized and OFDM modulated massive MIMO systems for an adjacent channel interference (ACI) scenario under frequency selective fading and channel estimation errors. It is also the first to analyze the performance of oversampling ADCs for such a scenario.

Other than the investigated scenario being different from the existing studies in the literature, the difference of this work compared to the aforementioned studies dealing with the analysis of quantized uplink massive MIMO systems with low-resolution ADCs are as follows. In \cite{ICC_own,TWC_own,VTC_wideband_own}, temporal oversampling and corresponding performance analysis is performed for an uplink massive MIMO system with one-bit ADCs in a single-carrier environment and flat fading, which results in a high-complexity receiver in terms of baseband signal processing, whereas our study considers an OFDM system under frequency selective channel with low-resolution ADCs (not only one-bit), whose receiver complexity is changing almost linearly with oversampling. Another study \cite{Mollen_wideband} analyzes massive MIMO structures with one-bit ADCs under frequency selective fading. In that work, quantization noise is regarded as an uncorrelated distortion in time and space, which fails to hold when oversampling is performed or when the number of users or noise variance is not high \cite{Li_perf_analysis,jacobsson_lin_prec}. However, it will be seen that our analysis takes into account the temporal and spatial correlation in the quantization distortion, which enables an accurate analysis. Moreover, \cite{Li_perf_analysis} provides an analysis for flat fading channels for massive MIMO structures with one-bit ADCs and provide a short section for the analysis of frequency selective channel case claiming that extension from flat fading channel case is straightforward. However, the sizes of the covariance matrices found for the quantized received signal for frequency selective case are $MN\times MN$, $M$ and $N$ being the number of antennas and block length (the number of samples in a coherence interval or in pilot duration). This large size makes their use in the performance analysis and channel estimation (covariance matrix inverse is used in channel estimation) infeasible in terms of computational complexity (even their storage in memory during simulations is problematic) unless the block length or number of antennas is very small, which is not the case for massive MIMO. There is a similar complexity problem in \cite{jacobsson_lin_prec}, where the matrix sizes involved in the calculation of the performance metrics are as large as $MN\times MN$. This problem is addressed in \cite{jacobsson_aggragate} by combining frequency domain operation with time domain operation for the calculation of necessary covariance matrices. However, many important points regarding the construction of the signal model using Bussgang decomposition or proofs regarding how to find correlation matrices in frequency domain from time domain matrices or vica versa are omitted. \color{black}Furthermore, the calculation of a quantization noise covariance matrix in \cite[Eqn.28]{jacobsson_aggragate} is valid only when the quantization noise is uncorrelated over the time dimension, which does not hold for very low-resolution ADCs. Although the quantization noise covariance matrix calculation of \cite{jacobsson_aggragate} is shown to provide accurate results in \cite{jacobsson_aggragate}, this is mostly due to the fact that the investigated ADC bit resolution in \cite{jacobsson_aggragate} (6 bits) is rather high. \color{black}The more general version taking into account the time domain correlation is provided in Proposition~\ref{prop:error_cov_freq} in this work.  Moreover, the effect of system parameters such as the number of receive antennas or oversampling rate cannot be deduced from the signal-to-interference-noise-and-distortion ratio (SINDR) expressions in \cite{jacobsson_lin_prec,jacobsson_aggragate}, whereas we provide some approximate expressions for SINDR in Proposition~\ref{prop:approx_SINDR}, in which the effect of system parameters can easily be followed. Moreover, the analysis in \cite{jacobsson_lin_prec,jacobsson_aggragate} is only performed for the perfect \color{black} channel state information \color{black}(CSI) case, whereas we propose a channel estimation algorithm and include the effect of imperfect CSI in this study.
In summary, the contribution items are as provided below.
\begin{itemize}
\item{This study is the first to analyze the performance of uplink massive MIMO systems with low-resolution ADCs in terms of error-rate and ergodic capacity under an ACI scenario. The analysis covers the frequency selective fading channel conditions. We obtain two types of analytical expressions for SINDR, one being more precise, whose accuracy is verified with simulations, and the other being less accurate but able to provide clear insights into the system performance and parameters. We show both analytically and with simulations that it is possible to combat ACI by increasing the number of receive antennas.}
\item{We analyze the effect of oversampling in such systems and show analytically that oversampling is also effective to suppress ACI. We also show that significant performance gains can be obtained by oversampling either with simulations or theoretical analysis.}
\item{We propose a linear minimum mean square error (LMMSE) based channel estimation algorithm taking into account the effect of an adjacent channel interferer. The provided analysis is able to incorporate the effect of imperfect CSI on the system performance.}
\item{We extend our analysis to multi-bit ADCs and discuss whether to increase the ADC resolution or oversampling rate by making comparisons in terms of error-rate performance while ADC power consumptions are kept constant.}
\item{The analysis in the paper is general in that it can also be applied to the scenario where no ACI is present. For no ACI case, the analysis in the paper}
\begin{itemize}
\item{requires much less memory resources than the ones in \cite{Li_perf_analysis,jacobsson_lin_prec} for SINDR calculations while providing closed-form expressions for quantization noise covariance matrix for one-bit ADCs as in (\ref{eqn:dist_simplified}) and (\ref{eqn:arcsine_rule}) (more details for this item are mentioned previously),}
\item{takes into account the temporal and spatial correlation for the quantization noise, which are neglected in \cite{Mollen_wideband,jacobsson_aggragate}, among which \cite{Mollen_wideband} does not cover the effect of oversampling,} 
\item{can clearly show the impact of system parameters (such as number of antennas, oversampling rate, etc.) and imperfect CSI on the system performance unlike \cite{jacobsson_aggragate, jacobsson_lin_prec}, where no channel estimation technique is proposed.}
\end{itemize}
\end{itemize}
A preliminary version for a subset of the mentioned contribution items is to appear as a conference paper \cite{Globecom_submitted_AIC}.

%In the conference version, we consider only the one-bit ADC case and mainly perfect receiver channel state information (CSI) case, although some preliminary results concerning imperfect CSI are revealed, without giving any details regarding how channel estimation is performed. Moreover, most of the proofs for the propositions presented in this paper are omitted in the conference version. Furthermore, SINDR expressions through which the effect of system parameters on system performance are only given for perfect CSI in the conference paper, whereas they are extended to the imperfect CSI case in this paper.

%In a nutshell, the contribution items for this paper are as follows:

%Moreover, proofs for some analytical points that have a similarity to the case where no interferer exists, which were assumed without any proof or discussion in \cite{jacobsson_aggragate} will also be provided.

\textit{Notation:} $c$ is a scalar, $\mathbf{c}$ is a column vector, $\mathbf{C}$ is a matrix and $\mathbf{C}^H$, $\mathbf{C}^T$, and $\mathbf{C}^*$ represent the Hermitian, transpose, and conjugate of matrix $\mathbf{C}$, respectively. $(\mathbf{C})_{m,n}$ stands for the element of matrix $\mathbf{C}$ at its $m^{th}$ row and $n^{th}$ column and $|\mathcal{G}|$ represents the cardinality of the set $\mathcal{G}$. $\mathbb{E}[.]$ takes the expectation of its operand.  $\Re(.)$ and $\Im(.)$ take the real and imaginary parts of their operands and $j=\sqrt{-1}$. $||.||$ corresponds to the Euclidean norm. $\mathbf{0}_K$ and $\mathbf{I}_K$ are zero and identity matrices with size $K\times K$; and $\mathrm{diag}(\mathbf{C})$ is the diagonal matrix, whose diagonal entries are equal to the diagonal entries of $\mathbf{C}$. Moreover, $\text{Tr}[.]$ is the trace operator. Furthermore, $\Phi(.)$ is the standard normal cumulative distribution function (CDF), that is, $\Phi(x)=\frac{1}{\sqrt{2\pi}}\int_{-\infty}^{x}e^{-z^2/2}dz$. 
%$\mathbbm{1}_\mathcal{A}(.)$ is the indicator function giving an output of 1 if the operand is an element of the set $\mathcal{A}$ and 0 otherwise.

\section{System Model}

\begin{figure}[t]
\centering
\includegraphics[width=0.9\columnwidth]{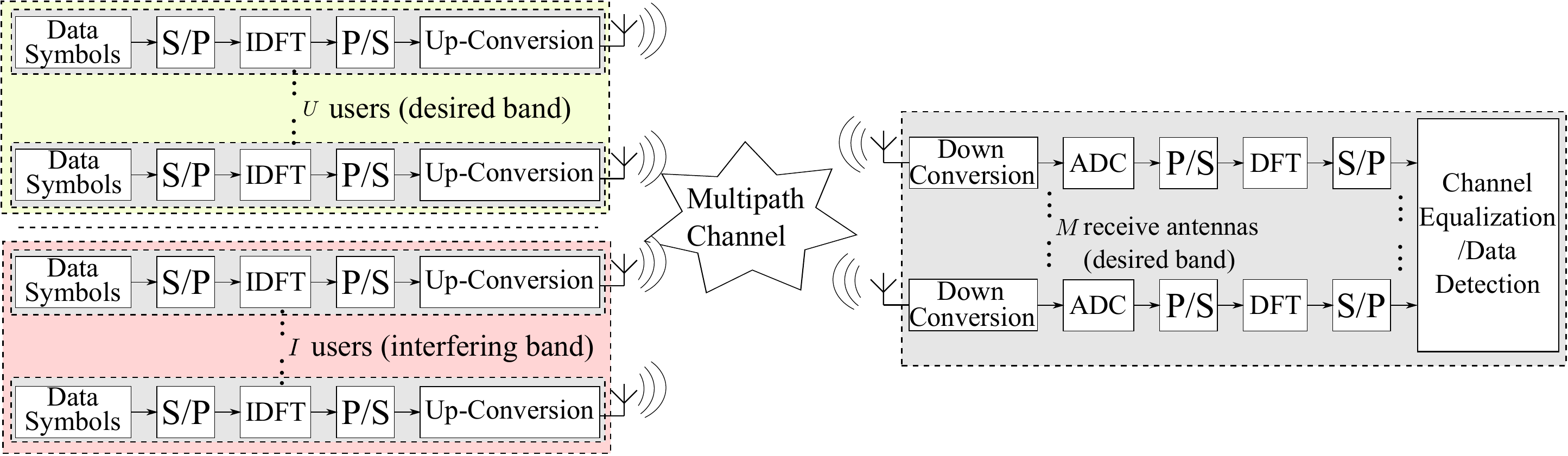}
\caption{Multi-user uplink massive MIMO-OFDM block diagram in an interfering band scenario.}
\label{fig:system_model}
\end{figure}

\begin{figure}[t]
\centering
\includegraphics[width=0.88\columnwidth]{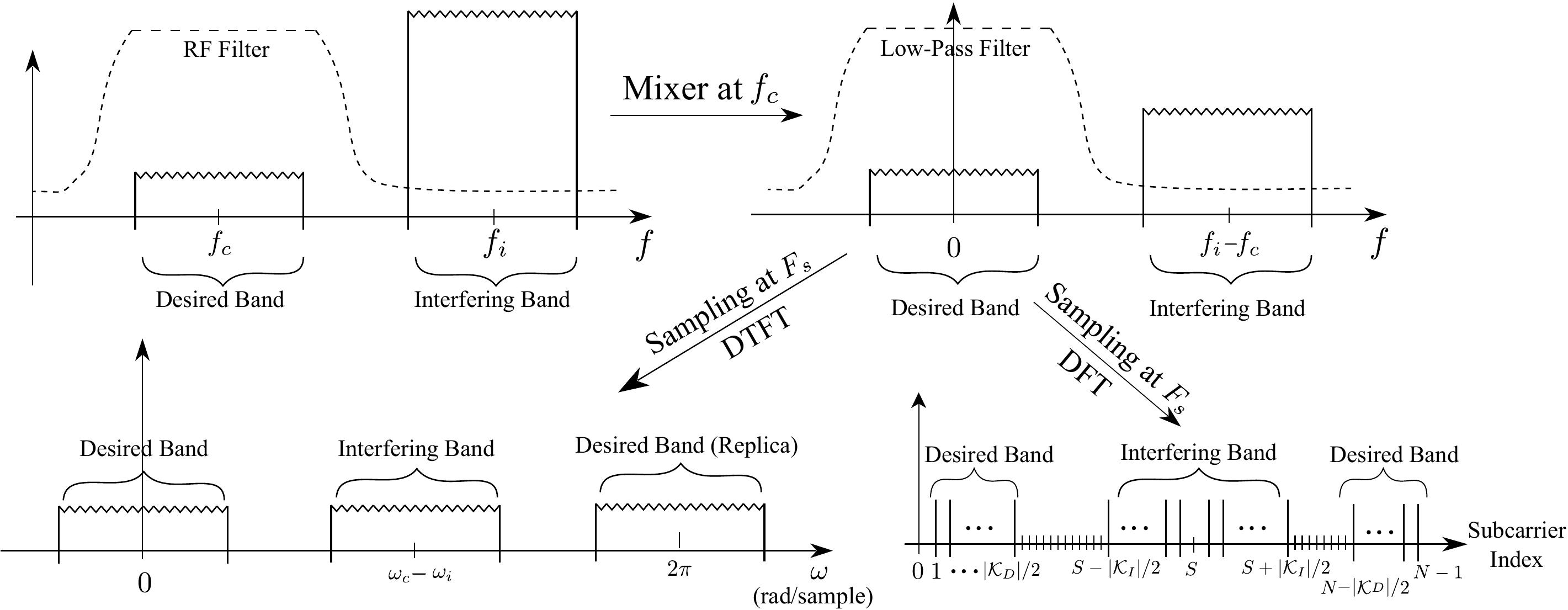}
\caption{Example plots for the spectrum of the signals at various receiver stages.}
\label{fig:spectrum_receiver}
\end{figure}
%On the bottom right, the subcarrier assignments for desired and interfering band users are shown. $S$ represents the subcarrier index for the center of the interfering band.
We consider the uplink scenario depicted in Fig.~\ref{fig:system_model}. It is assumed that $U$ users send their information to a base station in an OFDM massive MIMO setting through a set of subcarriers that are assigned to them. This set of subcarriers will be denoted by $\mathcal{K}_D$ and will be referred to as the $\textit{desired band}$ in this study. The desired band users are illustrated with green background in Fig.~\ref{fig:system_model}. The receiver side in Fig.~\ref{fig:system_model} is a typical OFDM receiver, in which the ADC block is assumed to have low-resolution in this study. Another group of $I$ users, whose assigned set of subcarriers is denoted by the set $\mathcal{K}_I$, is acting as an interfering source to the users in the desired band. These users are shaded with red background in Fig.~\ref{fig:system_model}. The set of subcarriers in $\mathcal{K}_I$ will also be referred to as \textit{interfering band} in the remainder of this study. It should be noted that although the interference is from an adjacent band, it may not be suppressed enough due to near/far effect despite all analog filters involved in the down-conversion stages. Example plots for the spectrum of the signals at various points of a zero intermediate frequency (IF) receiver are provided in Fig.~\ref{fig:spectrum_receiver}, where $f_c$ and $f_i$ are the carrier frequencies for the signals at the desired and interfering bands, respectively. The top left and right plots in Fig.~\ref{fig:spectrum_receiver} represent the  \color{black}power spectral densities \color{black}(PSD) of the signals at the \color{black}radio-frequency \color{black} (RF) front end (just before the bandpass RF filter centered at $f_c$, for which the interfering band signal is much stronger compared to the desired band signal) and at the mixer output (before the low-pass filter (LPF) in the down-converter), respectively. Such a scenario can be considered as a typical long-term evolution (LTE) case in which different users are assigned to rectangular areas of resource blocks or subbands \cite{lte_resource_alloc}. The discrete-time Fourier transform (DTFT) and discrete Fourier transform (DFT) of the sampled (but unquantized) signal are also shown at the bottom two plots in Fig.~\ref{fig:spectrum_receiver}, where $\omega_c=2\pi f_c/F_s$, $\omega_i=2\pi f_i/F_s$ and $S=N(f_i-f_c)/(2F_s)$, $F_s$ being the sampling rate. Moreover, it is also assumed that the receiver samples fast enough to cover both the desired and interfering band to avoid any interference due to aliasing from the interfering band to the desired band. This is to study the isolated spectral leakage effect from the interfering to desired band owing to the non-linearity due to low-resolution ADCs, as ACI due to aliasing will occur even when there is no non-linearity, which is not the focus of this work. The sets of users in the desired and interfering band are denoted by $\mathcal{U}_D=\{1,2,\ldots,U\}$ and $\mathcal{U}_I=\{U+1,U+2,\ldots,U+I\}$, respectively.

\subsection{Signal Model}
\label{sec:signal_model}

We denote the complex data symbol of a user $u$ transmitted at the $k^{th}$ subcarrier by $\tilde{s}_u[k]$, $k=0,1,\ldots,N-1$, where $N$ is the DFT size. Not all subcarriers are occupied, that is, $\tilde{s}_u[k]=0$ for $k\notin\mathcal{K}_D$ when $u\in\mathcal{U}_D$ or $k\notin\mathcal{K}_I$ when $u\in\mathcal{U}_I$, as shown in the bottom right of Fig.~\ref{fig:spectrum_receiver}. The oversampling rate for the users in desired and interfering bands, $\beta_D$ and $\beta_I$, are defined as a ratio of the total number of subcarriers $N$ to the number of occupied subcarriers, that is, $\beta_D\triangleq N/|\mathcal{K}_D|$ and $\beta_I\triangleq N/|\mathcal{K}_I|$. Increasing $N$ while $|\mathcal{K}_D|$ or $|\mathcal{K}_I|$ is fixed is termed as ``oversampling" since we consider the case that the OFDM symbol duration $NT_s$ is fixed, where $T_s$ is the sampling period, requiring that the sampling rate ($1/T_s$) is increased while $N$ is increased. This also ensures that the transmission bandwidth of the desired channel users is kept the same when the oversampling rate $\beta_D$ is increased, as the subcarrier spacing $1/(NT_s)$ and the number of occupied subcarriers $|\mathcal{K}_D|$ are fixed, which results in a fixed transmission bandwidth of $|\mathcal{K}_D|/(NT_s)$.

Following those definitions, the discrete-time signal of the $u^{th}$ user at the inverse DFT (IDFT) output, $s_u[n]$, can be expressed as
\begin{equation}
s_u[n]=
  \begin{cases}
       \dfrac{\rho_d}{\sqrt{N}}\sum_{k\in\mathcal{K}_D}\tilde{s}_u[k]e^{j2\pi nk/N} & \text{if } u\in\mathcal{U}_D, \\
 \dfrac{\rho_i}{\sqrt{N}}\sum_{k\in\mathcal{K}_I}\tilde{s}_u[k]e^{j2\pi nk/N} & \text{if } u\in\mathcal{U}_I,
  \end{cases}
\label{eqn:IDFT_op_sig}
\end{equation}
for $n=0,1,\ldots,N-1$. \color{black} Here $\rho_d$ and $\rho_i$ are the average transmit power parameters for the desired or interfering band users. \color{black} Moreover, data symbols have unit energy, that is, $\mathbb{E}[|\tilde{s}_u[k]|^2]=1$. For simple equalization at the receiver side, a \color{black}cyclic prefix \color{black}(CP) of length $L_{cp}$ is added to the beginning of the OFDM symbol such that $s_u[n]=s_u[N+n]$ for $n=-L_{cp}+1,\ldots,-1$. It is required that $L_{cp}\geq L-1$, $L$ being the number of channel taps\footnote{$L$ will be increased when the sampling rate ($1/T_s$) is increased as the delay spread of the channels does not change with the sampling rate.}.

After IDFT, the up-conversion block forms the OFDM modulated continuous time signal at the carrier frequency, whose baseband equivalent expression for user $u$ is
\begin{equation}
\label{eqn:tx_sig}
s_u(t)=\sum_{n=-L+1}^{N-1}s_u[n]p(t-nT_s),
\end{equation}
where $p(t)=\sinc(t/T_s)$ and $-L_{cp}T_s<t<NT_s$. In (\ref{eqn:tx_sig}), the effect of the neighboring OFDM frames (blocks) is omitted, as inter-block interference will not be present when CP is discarded at the receiver. Although the bandwidth of $p(t)$ seems to increase with the sampling rate, the bandwidth of $s_u(t)$ is limited to $|\mathcal{K}_D|/(NT_s)$ or $|\mathcal{K}_I|/(NT_s)$ due to the correlation statistics of $s_u[n]$ in time.
%It seems from (\ref{eqn:tx_sig}) that the carrier frequencies for desired and interfering band users are the same. However, the subcarrier assignments according to the disjoint sets $\mathcal{K}_D$ and $\mathcal{K}_I$ make sure that the two bands do not overlap in the frequency domain (see an example illustration for the desired and interfering bands in the bottom right of Fig.~\ref{fig:system_model}, where the parameter $S$ is defined in the figure caption).
The baseband equivalent received signal at the $m^{th}$ antenna can be expressed as
\begin{equation}
\label{eqn:rec_sig}
r_m(t)=\sum_{u=1}^{U+I} h'_{m,u}(t)\ast s_u(t) + w_m(t),
\end{equation}
where $\ast$ represents the convolution operation and $w_m(t)$ is the filtered additive white complex Gaussian noise process at the $m^{th}$ antenna. Moreover, $h'_{m,u}(t)$ is the continuous-time channel impulse response between the $u^{th}$ user and the $m^{th}$ antenna. In writing (\ref{eqn:rec_sig}), it is assumed that the bandpass RF filter and the LPF after the mixer have a flat response over the frequency bands $[f_c-F_s/2, \ f_c+F_s/2]$ and $[-F_s/2,\ F_s/2]$, respectively. Assuming that the sampling rate of the receiver is the same as that of the transmitter, (\ref{eqn:tx_sig}) and (\ref{eqn:rec_sig}) imply that the discrete-time received signal at the $m^{th}$ antenna, namely $r_m[n]=r_m(nT_s)$, can be expressed as follows:
\begin{equation}
\label{eqn:rec_sig_discr_time}
r_m[n]=r_m(nT_s)=\sum_{u=1}^{U+I}\sum_{\ell=0}^{L-1}h_{m,u}[\ell]s_u[n-\ell] + w_m[n],
\end{equation}
where $w_m[n]=w_m(nT_s)$, $h_{m,u}[\ell]=h_{m,u}(\ell T_s)$, in which $h_{m,u}(t)\triangleq h'_{m,u}(t)\ast \sinc(t/T_s)$. We assume that the channel coefficients $h_{m,u}[\ell]$ have complex Gaussian distribution and are uncorrelated, that is, $\mathbb{E}[h_{m,u}[\ell_1]h_{m,u}[\ell_2]^*]=p[\ell_1]\delta[\ell_1-\ell_2]$, as in the related studies \cite{Mollen_wideband,jacobsson_aggragate,jacobsson_lin_prec}. Here, $p[\ell]$ is the power delay profile of the channel satisfying $\sum_{l=0}^{L-1}p[\ell]=1$.
%where $w_m[n]=w_m(nT_s)$ and $h_{m,u}[\ell]=g_{m,u}(\ell T_s)$, $g_{m,u}(t)\triangleq h_{m,u}(t)\ast \sinc(t/T_s)$. For the wide-sense stationary uncorrelated channel scattering model (WSSUS) \cite{Goldsmith_WC}, $\mathbb{E}[h_{m,u}(\tau_1)h_{m,u}(\tau_2)^*]=\sigma_h^2\delta(\tau_1-\tau_2)$, where $\delta(\tau_1-\tau_2)$ is the continuous time impulse function, it follows that the channel taps are uncorrelated, that is, $\mathbb{E}[h_{m,u}[\ell_1]h_{m,u}[\ell_2]^*]=\delta[\ell_1-\ell_2]$. This is owing to the fact that $\mathbb{E}[h_{m,u}[\ell_1]h_{m,u}[\ell_2]^*]=\sigma_h^2R_p[\ell_1-\ell_2]$ where $R_p[\ell]\triangleq\int_{-\infty}^{\infty}p(t+\ell T_s)p^*(t)dt=\delta[\ell]$ for $p(t)=\sinc(t/T_s)$, $\delta[\ell]$ being the Kronecker delta function. 
%Moreover, the other assumption that $w_m[n]$ is white is valid under the assumptions that the RF bandpass filter has flat response over a bandwidth of $F_s$ around $F_c$ and zero gain in the stopband, in addition to the assumption that LPF after the mixer has a flat response over a bandwidth of $F_s/2$ in the pass band and zero gain in the stopband. The variance of the noise process is denoted by $N_o$.
A more compact version of (\ref{eqn:rec_sig_discr_time}) is
\begin{equation}
\label{eqn:rec_sig_discr_time_mtx_vec}
\mathbf{r}[n]=\sum_{\ell=0}^{L-1}\mathbf{H}[\ell]\mathbf{s}[n-\ell] + \mathbf{w}[n],
\end{equation}
where $\mathbf{H}[\ell]$ is an $M\times (U+I)$ matrix whose element at the $m^{th}$ row and the $u^{th}$ column is $h_{m,u}[l]$. The elements of $\mathbf{H}[\ell]$ are also assumed to be uncorrelated. Moreover, $\mathbf{s}[n]$ is a column vector whose $u^{th}$ element is $s_u[n]$. Furthermore, $\mathbf{w}[n]$ and $\mathbf{r}[n]$ are column vectors whose $m^{th}$ element is equal to $w_m[n]$ and $r_m[n]$, respectively.

The signal at the output of the one-bit quantizer (the case for the higher bit resolutions is presented in Section~\ref{sec:higher_res_ADC}), namely $\mathbf{d}[n]$, can be written in terms of its input, $\mathbf{r}[n]$ as
\begin{equation}
\label{eqn:quantized signal}
\mathbf{d}[n]=\mathrm{sign}(\Re(\mathbf{r}[n]))+j\mathrm{sign}(\Im(\mathbf{r}[n])),
\end{equation}
where $\mathrm{sign}(.)$ is the $\mathrm{sign}$ function. The DFT of the quantizer output is also taken to obtain the input signal to the channel equalization and data detection block, namely $\mathbf{\tilde{d}}[k]$, as follows:
\begin{equation}
\label{eqn:signal_after_fft}
\tilde{\mathbf{d}}[k]=\sum_{n=0}^{N-1}\mathbf{d}[n]e^{-j2\pi nk/N}.
\end{equation}
How to obtain the data estimates based on $\mathbf{\tilde{d}}[k]$ will be considered in the next section.

\section{Performance Analysis}
\label{sec:perf_analysis}

%In this section, the performance analysis of one-bit quantized massive MIMO system with an adjacent interferer scenario, whose details are given in Section \ref{sec:signal_model} will be presented. The performance metrics to be investigated will be bit error rate and ergodic capacity.

For a tractable analysis of the non-linear system with one-bit ADCs, the Bussgang decomposition \cite{Li_perf_analysis}, which enables a linear input-output relation for a non-linear system, will be employed. Before that, it is necessary to reexpress (\ref{eqn:rec_sig_discr_time}) as
\begin{equation}
\underline{\mathbf{r}}=\underline{\mathbf{H}}\hspace{2pt}\underline{\mathbf{s}}+\underline{\mathbf{w}},
\end{equation}
where $\underline{\mathbf{r}}=\left[\mathbf{r}\left[N-1\right]^T\ \mathbf{r}\left[N-2\right]^T \ \cdots \ \mathbf{r}\left[0\right]^T\right]^T$, \hspace{10pt} $\underline{\mathbf{s}}=\left[\mathbf{s}\left[N-1\right]^T \ \mathbf{s}\left[N-2\right]^T \ \cdots \ \mathbf{s}\left[0\right]^T\right]^T$, \hspace{10pt} $\underline{\mathbf{w}}=\left[\mathbf{w}\left[N-1\right]^T \ \mathbf{w}\left[N-2\right]^T \ \cdots \ \mathbf{w}\left[0\right]^T\right]^T$, and $\underline{\mathbf{H}}$ is a block circulant matrix of size $NM\times N(U+I)$ that can be expressed as follows:
\begin{align}
\underline{\mathbf{H}}
\begin{split}
=\left[ {\begin{array}{ccccccc} \mathbf{H}[0] & \mathbf{H}[1] & \cdots & \mathbf{H}[L-1] & \mathbf{0} & \cdots & \mathbf{0}\\
\mathbf{0} & \mathbf{H}[0] & \mathbf{H}[1] & \ddots & \ddots & \ddots & \vdots\\
\vdots & \vdots & \ddots & \ddots & \ddots & \ddots & \mathbf{0} \\
\mathbf{0} & \cdots & \mathbf{0} & \mathbf{H}[0] & \cdots & \cdots & \mathbf{H}[L-1]\\
\vdots & \vdots & \ddots & \ddots & \ddots & \ddots & \cdots\\
\mathbf{H}[2] & \mathbf{H}[1] & \cdots & \mathbf{H}[L-2] & \cdots & \mathbf{H}[0] & \mathbf{H}[1]\\ 
\mathbf{H}[1] & \mathbf{H}[2] & \cdots & \mathbf{H}[L-1] & \cdots & 0 & \mathbf{H}[0]\\ 
\end{array} } \right]
\end{split}\label{big_chan_mtx}
\end{align}
This is the MIMO extension of the circulant channel matrix defined for a single-input single-output (SISO) OFDM scenario in \cite{Goldsmith_WC}. According to the Bussgang decomposition \cite{Li_perf_analysis,rowe1982memoryless},
\begin{equation}
\label{eqn:bussgang_lin}
\underline{\mathbf{d}}=\underline{\mathbf{A}}\hspace{2pt}\underline{\mathbf{r}}+\underline{\mathbf{q}},
\end{equation}
where $\underline{\mathbf{q}}$ is the equivalent quantization noise vector, $\underline{\mathbf{d}}=\left[\mathbf{d}\left[N-1\right]^T \ \mathbf{d}\left[N-2\right]^T \ \cdots \ \mathbf{d}\left[0\right]^T\right]^T$, and $\underline{\mathbf{A}}=\mathbf{C}^H_{\underline{\mathbf{d}}\hspace{1.5pt}\underline{\mathbf{r}}}\mathbf{C}_{\underline{\mathbf{r}}}^{-1}$, where $\mathbf{C}_{\underline{\mathbf{r}}}=\mathbb{E}[\underline{\mathbf{r}}\hspace{2pt}\underline{\mathbf{r}}^H]$ and $\mathbf{C}_{\underline{\mathbf{d}}\hspace{2pt}\underline{\mathbf{r}}^H}=\mathbb{E}[\underline{\mathbf{d}}\hspace{2pt}\underline{\mathbf{r}}^H]$. The size of $\underline{\mathbf{A}}$ is $NM\times NM$. This choice of $\underline{\mathbf{A}}$ minimizes the variance of the quantizer noise or equivalently makes $\underline{\mathbf{r}}$ uncorrelated with $\underline{\mathbf{q}}$. For a one-bit quantizer, assuming zero-mean Gaussian inputs\footnote{\label{ftnt:Quant_inp_gauss}This assumption is approximately true even when the transmitted symbols are from a finite cardinality set rather than being Gaussian distributed. The ADC input at the $m^{th}$ antenna can be written as sum of $\mathcal{S}_U^m$ and $\mathcal{S}_I^m$, where $\mathcal{S}_U^m$ is the sum of $U|\mathcal{K_D}|L$ independent identically distributed (i.i.d.) signals with finite variance from the desired band, and $\mathcal{S}_I^m$ is the sum of $I|\mathcal{K_I}|L$ i.i.d. signals from the interfering band. Due to the central limit theorem, $\mathcal{S}_U^m$ and $\mathcal{S}_I^m$ converge to Gaussian as $U|\mathcal{K_D}|L$ and $I|\mathcal{K_I}|L$ grow large, so does the ADC input $\mathcal{S}_U^m+\mathcal{S}_I^m$. \color{black} It can be shown by the Berry-Essen inequality that the difference between the CDFs of $\mathcal{S}_U^m$ or $\mathcal{S}_I^m$ and the CDF of a Gaussian random variable with the same mean and variance is always less than 0.02, even when $U$ and $I$ are as low as $4$, $L=3$ and $|\mathcal{K_D}|$ or $|\mathcal{K_I}|$ is $128$. Since CDFs are between 0 and 1, an error of 0.02 is negligable. For higher $I|\mathcal{K_I}|L$ or $U|\mathcal{K_D}|L$, this error will be much less (error decreases with $\sqrt{U|\mathcal{K_D}|L}$ or $\sqrt{I|\mathcal{K_I}|L}$).\color{black}}, the following holds \cite{Li_perf_analysis}:
%\setul{1pt}{0.2pt}
%\begin{equation}
%\label{eqn:correlation_mtx_exp}
%\mathbf{C}_{\text{\ul{$\mathbf{d}$}}\hspace{2pt}\text{\ul{$\mathbf{r}$}}^H}=\sqrt{\dfrac{4}{\pi}}\mathbf{C}_{\text{\ul{$\mathbf{r}$}}}\mathrm{diag}(\mathbf{C}_{\text{\ul{$\mathbf{r}$}}})^{-0.5}.
%\end{equation}
%By substituting (\ref{eqn:correlation_mtx_exp}) into (\ref{eqn:linear_gain_MMSE}), it can be found that
\begin{equation}
\label{eqn:a_mtx_exp}
\underline{\mathbf{A}}=\sqrt{\dfrac{4}{\pi}}\mathrm{diag}(\mathbf{C}_{\text{\ul{$\mathbf{r}$}}})^{-0.5}=\sqrt{\dfrac{4}{\pi}}\mathrm{diag}\left(\underline{\mathbf{H}}\mathbf{R_{\underline{s}}}\underline{\mathbf{H}}^H+N_o\mathbf{I}\right)^{-0.5},
\end{equation}
where $\mathbf{R_{\underline{s}}}=\mathbb{E}[\mathbf{\underline{s}}\mathbf{\underline{s}}^H]$. Although a closed form expression as in (\ref{eqn:a_mtx_exp}) is available to calculate the matrix $\underline{\mathbf{A}}$, its calculation is not simple, whose complexity is in the order of $N^3M^2$, which is very large for a typical massive MIMO scenario. Therefore, an alternative low-complexity approach will be presented. Since $\underline{\mathbf{A}}$ is a diagonal matrix, (\ref{eqn:bussgang_lin}) can be modified as
\begin{equation}
\label{eqn:bussgang_lin_short_time_ind}
\mathbf{d}[n]=\mathbf{A}[n]\mathbf{r}[n]+\mathbf{q}[n],
\end{equation}
for $n=1,\ldots,N$, where $\mathbf{A}[n]$ is a diagonal matrix whose diagonal elements are equal to a set of diagonal elements of $\underline{\mathbf{A}}$, which are, $\left[(\underline{\mathbf{A}})_{n,n} (\underline{\mathbf{A}})_{n+1,n+1} \cdots \ (\underline{\mathbf{A}})_{n+M-1,n+M-1} \right]$. Moreover, as $\mathbf{r}[n]$ is stationary owing to the addition of the cyclic prefix and $\mathbf{A}[n]$ is diagonal, it can be shown that 
$\mathbf{A}[n]=\mathbf{A}[n']\triangleq \mathbf{A}, \forall n,n'$, thus
%
%according to (\ref{eqn:a_mtx_exp}) and (\ref{eqn:bussgang_lin_short_time_ind}), the following equality holds:
%\begin{equation}
%\label{eqn:a_mtx_small_piece}
%\mathbf{A}[n]=\sqrt{\dfrac{4}{\pi}}\mathrm{diag}\left(\mathbf{C}_{\mathbf{r}[n]}\right),
%\end{equation} 
%where $\mathbf{C}_{\mathbf{r}[n]}\triangleq\mathbb{E}\left[\mathbf{r}[n]\mathbf{r}[n]^H\right]$. Since $\mathbf{r}[n]$ in (\ref{eqn:rec_sig_discr_time_mtx_vec}) can be shown to be stationary, defining $\mathbf{C_{r}}[m]\triangleq\mathbb{E}\left[\mathbf{r}[n]\mathbf{r}[n-m]^H\right]$, $\mathbf{C}_{\mathbf{r}[n]}=\mathbf{C_{r}}[0]$. This implies that the right hand side of (\ref{eqn:a_mtx_small_piece}) does not depend on the time index $n$ which implies $\mathbf{A}[n]=\mathbf{A}[n']\triangleq \mathbf{A}, \forall n,n'$,
\begin{equation}
\label{eqn:no_time_index_quant_sig}
\mathbf{d}[n]=\mathbf{A}\mathbf{r}[n]+\mathbf{q}[n],
\end{equation}
where $\mathbf{A}=\sqrt{4/\pi}\mathrm{diag}\left(\mathbf{C_r}[0]\right)^{-0.5}$ is an $M\times M$ matrix, in which $\mathbf{C_{{r}}}[m]\triangleq\mathbb{E}\left[\mathbf{{r}}[n]\mathbf{{r}}[n-m]^H\right]$. It is still hard to find $\mathbf{C_r}[0]$ in the time domain using (\ref{eqn:rec_sig_discr_time_mtx_vec}) as there exist correlation between the time domain symbols $\mathbf{s}[n]$ due to oversampling. It will also be more convenient to work in the frequency domain as the final detection of the data symbols will be performed in that domain, thus any SINDR expression to be used in the analysis should be found for the frequency domain observations.  Therefore, the analysis continues by taking the DFT of both sides of (\ref{eqn:no_time_index_quant_sig}), yielding
\begin{equation}
\label{eqn:no_time_index_quant_sig_freq_dom}
\mathbf{\tilde{d}}[k]=\mathbf{A}\mathbf{\tilde{r}}[k]+\mathbf{\tilde{q}}[k],
\end{equation}
where $\tilde{\mathbf{r}}[k]$ has a simple expression that can be found by using (\ref{eqn:rec_sig_discr_time_mtx_vec}) and considering the circulant property of the channel convolution matrix due to the addition of the cyclic prefix as
\begin{equation}
\label{eqn:freq_dom_eq}
\tilde{\mathbf{r}}[k]=\begin{cases}
\rho_d\sqrt{N}\tilde{\mathbf{H}}[k]\tilde{\mathbf{s}}[k]+\tilde{\mathbf{w}}[k], & \text{if } k\in\mathcal{K}_D, \\
\rho_i\sqrt{N}\tilde{\mathbf{H}}[k]\tilde{\mathbf{s}}[k]+\tilde{\mathbf{w}}[k], & \text{if } k\in\mathcal{K}_I, \\
0 & \text{otherwise},
\end{cases}
\end{equation}
where $\tilde{\mathbf{s}}[k]\triangleq\left[\tilde{s}_1[k]\ \tilde{s}_2[k] \ \cdots \ \tilde{s}_{U+I}[k] \right]$, $k=0,1,\ldots,N-1$. Moreover, $\mathbf{\tilde{r}}[k]=\sum_{n=0}^{N-1}\mathbf{r}[n]e^{-j2\pi nk/N}$, $\mathbf{\tilde{H}}[k]=\sum_{n=0}^{N-1}\mathbf{H}[n]e^{-j2\pi nk/N}$, $\mathbf{\tilde{w}}[k]=\sum_{n=0}^{N-1}\mathbf{w}[n]e^{-j2\pi nk/N}$. Next, we define $\mathbf{C}_{\mathbf{\tilde{r}}[k]}\triangleq\mathbb{E}\left[\mathbf{\tilde{r}}[k]\mathbf{\tilde{r}}[k]^H\right]$. Since $\mathbb{E}[\mathbf{\tilde{r}}[k]\mathbf{\tilde{r}}[k']^H]=\mathbf{0}$ for $k\neq k'$, $\mathbf{C_{r}}[m]$ can be found according to the following proposition.
\begin{prop} \label{prop:time_dom_correlation} $\mathbf{C_{r}}[m]$ \text{can be computed from} $\mathbf{C}_{\mathbf{\tilde{r}}[k]}$ as 
\begin{equation}
\mathbf{C_{r}}[m]=\dfrac{1}{N^2}\sum_{k=0}^{N-1}\mathbf{C}_{\mathbf{\tilde{r}}[k]}e^{j2\pi mk/N}.
\end{equation}
\begin{proof}
See Appendix \ref{apdx:prop1_proof}.
\end{proof}
\end{prop}
$\mathbf{C}_{\mathbf{r}}[m]$ can be calculated by taking the IDFT of $\mathbf{C}_{\mathbf{\tilde{r}}[k]}$, which can be found using (\ref{eqn:freq_dom_eq}) as
\begin{equation}
\label{eqn:freq_dom_corr}
\mathbf{C}_{\mathbf{\tilde{r}}[k]}=\begin{cases}
\rho_d^2N\tilde{\mathbf{H}}[k]\tilde{\mathbf{H}}[k]^H+NN_o\mathbf{I}, & \text{if } k\in\mathcal{K}_D, \\

\rho_i^2N\tilde{\mathbf{H}}[k]\tilde{\mathbf{H}}[k]^H+NN_o\mathbf{I}, & \text{if } k\in\mathcal{K}_I, \\
0, & \text{otherwise}.
\end{cases}
\end{equation}
%As the size of matrix $\tilde{\mathbf{H}}[k]$ is independent of $N$, the calculation of $\mathbf{C}_{\mathbf{\tilde{r}}[k]}$ grows linearly with $U+I$. The calculation of $\mathbf{C}_r[0]$, which will yield $\mathbf{A}$, using (\ref{eqn:prop_prf_l4}) also require $N$ complex multiplications, which means the total complexity of the calculation of $\mathbf{C}_r[0]$ or $\mathbf{A}$ grows linearly with $N$. Moreover, it will also be shown that $\mathbf{C_r}[m]$ found in Proposition \ref{prop:ifft_cov_mtx} will also be used to find the covariance matrix of quantizer distortion $\mathbf{q}[k]$, namely $\mathbf{C}_{\mathbf{q}[k]}=\mathbb{E}[\mathbf{q}[k]\mathbf{q}[k]^H]$ in (\ref{eqn:no_time_index_quant_sig_freq_dom}). 
What remains is to find the covariance matrix of the quantization distortion $\mathbf{\tilde{q}}[k]$, namely $\mathbf{C}_{\mathbf{\tilde{q}}[k]}=\mathbb{E}[\mathbf{\tilde{q}}[k]\mathbf{\tilde{q}}[k]^H]$. Consider the quantization noise vector $\underline{\mathbf{q}}$ in (\ref{eqn:bussgang_lin}). $\mathbf{C}_{\underline{\mathbf{q}}}=\mathbb{E}[\underline{\mathbf{q}} \ \underline{\mathbf{q}}^H]$ is given by
\begin{equation}
\label{eqn:quant_dist_long_cov}
\mathbf{C}_{\underline{\mathbf{q}}}=\mathbf{C}_{\underline{\mathbf{d}}}-\underline{\mathbf{A}}\mathbf{C}_{\underline{\mathbf{r}}}\underline{\mathbf{A}}^H.
\end{equation}
The matrix sizes in (\ref{eqn:quant_dist_long_cov}) are $MN\times MN$, which can be very large, and require vast amount of memory resources for the computations (for instance, with $M=64, N=1024$, each matrix in (\ref{eqn:quant_dist_long_cov}) requires about 32 GB space for double precision number format). Such large matrices are also present in the studies \cite{Li_perf_analysis,jacobsson_lin_prec}. Therefore, an alternative method will be proposed to work with matrices of feasible sizes. From the block Toeplitz structure of $\mathbf{C}_{\underline{\mathbf{q}}}$ and the fact that $\underline{\mathbf{A}}$ is a diagonal matrix, it can be shown that
\begin{align}
\label{eqn:dist_simplified}
\mathbf{C_q}[m]=\mathbf{C_d}[m]-\mathbf{A}\mathbf{C_r}[m]\mathbf{A}^H,
\end{align}
%\hspace{500pt}
\begin{equation}
\begin{split}
\label{eqn:arcsine_rule}
\mathbf{C_d}[m]&=\dfrac{4}{\pi}\left(\mathrm{asin}\left(\mathbf{D_r}[m]^{-\frac{1}{2}}\Re(\mathbf{C_r}[m])\mathbf{D_r}[m]^{-\frac{1}{2}}\right)+j\mathrm{asin}\left(\mathbf{D_r}[m]^{-\frac{1}{2}}\Im(\mathbf{C_r}[m])\mathbf{D_r}[m]^{-\frac{1}{2}}\right)\right),
\end{split}
\end{equation}
where $\mathbf{D_r}[m]=\mathrm{diag}\left(\mathbf{C_r}[m]\right)$ and $m=0,1,\ldots,N-1$. (\ref{eqn:arcsine_rule}) is a result of the \textit{arcsine law} \cite{van1966spectrum}. Note that the matrix sizes in (\ref{eqn:dist_simplified}) and (\ref{eqn:arcsine_rule}) are $M\times M$ and the memory requirement to hold the matrices in (\ref{eqn:dist_simplified}) is $N$ times smaller compared to (\ref{eqn:quant_dist_long_cov}). Moreover, it can be noted that the temporal and spatial correlation of quantization noise is taken into account since $\mathbf{C_q}[m]$ can be calculated for $m\neq0$ using (\ref{eqn:dist_simplified}) and is not necessarily a diagonal matrix. Now that every matrix in (\ref{eqn:dist_simplified}) is known, $\mathbf{C_q}[m]$ can be calculated. The next step is to find $\mathbf{C}_{\mathbf{\tilde{q}}[k]}=\mathbb{E}[\mathbf{\tilde{q}}[k]\mathbf{\tilde{q}}[k]^H]$ from $\mathbf{C_q}[m]$, which is performed in the following proposition.
\begin{prop} \label{prop:error_cov_freq} $\mathbf{C}_{\mathbf{q}[k]}$ \text{can be computed from} $\mathbf{C_q}[m]$ as
\begin{equation}
\mathbf{C}_{\mathbf{q}[k]}=\mathrm{DFT}_{N,k}\left\lbrace\mathbf{\Gamma}[m]\right\rbrace+\mathrm{DFT}_{N,k}\left\lbrace\mathbf{\Gamma}[m]\right\rbrace^*-N\mathbf{C_{\tilde{q}}}[0],
\end{equation}
where $\mathbf{\Gamma}[m]\triangleq(N-m)\mathbf{C_q}[m]$ and $\mathrm{DFT}_{N,k}\left\lbrace\mathbf{\Gamma}[m]\right\rbrace=\sum_{m=0}^{N-1}\mathbf{\Gamma}[m]e^{-j2\pi mk/N}$.
\begin{proof}
%The proof is omitted due to space restrictions. The critical points in the proof are the utilization of the stationarity of $\mathbf{q}[n]$, DFT properties and some algebraic manipulations.
See Appendix \ref{apdx:prop2_proof}.
\end{proof}
%\begin{proof}
%\begin{align}
%    \mathbf{C}_{\mathbf{q}}&[k]=\mathbb{E}\left[\mathbf{q}[k]\mathbf{q}[k]^H\right] \label{eqn:prop_prf_l1}\\
%    &=\left[\dfrac{1}{N}\sum_{k=0}^{N-1}\sum_{k'=0}^{N-1}\mathbb{E}\left[\tilde{\mathbf{r}}[k]\tilde{\mathbf{r}}[k']^H\right]e^{j2\pi nk/N}e^{-j2\pi (n-m)k'/N}\right]\label{eqn:prop_prf_l2}\\
%    &=\left[\dfrac{1}{N}\sum_{k=0}^{N-1}\mathbb{E}\left[\tilde{\mathbf{r}}[k]\tilde{\mathbf{r}}[k]^H\right]e^{j2\pi nk/N}e^{-j2\pi (n-m)k/N}\right]\label{eqn:prop_prf_l3}\\
%    &=\dfrac{1}{N}\sum_{k=0}^{N-1}\mathbf{C}_{\mathbf{\tilde{r}}[k]}e^{j2\pi mk/N}\label{eqn:prop_prf_l4},
%\end{align}
%where (\ref{eqn:prop_prf_l1}) and (\ref{eqn:prop_prf_l2}) holds by definition, (\ref{eqn:prop_prf_l3}) is due to the fact that $\mathbb{E}\left[\tilde{\mathbf{r}}[k]\tilde{\mathbf{r}}[k']^H\right]=\mathbf{0}$ for $k\neq k'$ as the data symbols of the users are independent and (\ref{eqn:prop_prf_l4}) also holds by definition.
%\end{proof}
\end{prop}

\subsection{Data Detection}
For data detection, zero-forcing (ZF) combining is applied to the DFT output $\tilde{\mathbf{d}}[k]$ to obtain data estimates $\mathbf{\hat{x}}[k]$ as
\begin{equation}
\label{eqn:ZF_detection}
\mathbf{{\hat{x}}}[k]=\mathbf{\hat{B}}[k]{\mathbf{\tilde{d}}[k]},
\end{equation}
for $k\in\mathcal{K}_D$. In (\ref{eqn:ZF_detection}), $\mathbf{\hat{B}}[k]=\left(\mathbf{{\hat{H}}}[k]^H\mathbf{{\hat{H}}}[k]\right)^{-1}\mathbf{{\hat{H}}}[k]^H$, where $\mathbf{\hat{H}}[k]$ is the estimate for $\mathbf{\tilde{H}}[k]$. The details about the channel estimation is provided in Section~\ref{sec:chan_est}.
%To begin with the performance analysis, we first define the ZF combining matrix for the perfect CSI case as $\mathbf{\tilde{B}}[k]=\left(\mathbf{{\tilde{H}}}[k]^H\mathbf{{\tilde{H}}}[k]\right)^{-1}\mathbf{{\tilde{H}}}[k]^H$. 
Using (\ref{eqn:no_time_index_quant_sig_freq_dom}), (\ref{eqn:freq_dom_eq}) and (\ref{eqn:ZF_detection}), the $u^{th}$ element of $\mathbf{{\hat{x}}}[k]$, namely ${{\hat{x}}}_u[k]$, can be found as
\begin{equation}
\label{eqn:rec_sig_compnts}
{{\hat{x}}}_u[k]=g_u[k]+i_u[k]+n_u[k]+q_u[k],
\end{equation}
where
\begin{align}
g_u[k]&=g_u[k]'\tilde{s}_u[k],\ n_u[k]=\mathbf{{\hat{b}}}_u[k]^H\mathbf{A}\tilde{\mathbf{w}}[k],q_u[k]=\mathbf{{\hat{b}}}_u[k]^H{\mathbf{q}[k]}, \label{eqn:rec_sig_compnts1}\\
i_u[k]&=\rho_d\sqrt{N}\left[\sum_{z\neq u, z\in\mathcal{U}_d}\mathbf{{\hat{b}}}_u[k]^H\mathbf{A}{\mathbf{\hat{h}}_z}[k]\tilde{s}_z[k]+\sum_{z\neq u,z\in\mathcal{U}_d}{\mathbf{\hat{b}}_u}[k]^H\mathbf{A}\mathbf{\tilde{e}}_z[k]\tilde{s}_z[k]\right], \label{eqn:rec_sig_compnts3}
\end{align}
where $g_u[k]'=\rho_d\sqrt{N}\mathbf{\hat{b}}_u[k]^H\mathbf{A}{\mathbf{\tilde{h}}_u}[k]$. In (\ref{eqn:rec_sig_compnts}), $g_u[k]$ corresponds to the signal part, whereas, $i_u[k]$ term in (\ref{eqn:rec_sig_compnts}) contains the interference from other users (the left summation in the $i_u[k]$ expression) and the distortion caused by imperfect CSI (the right summation in the $i_u[k]$ expression). Furthermore, $n_u[k]$ and $q_u[k]$ correspond to the distortion caused by thermal noise and quantization, respectively. Moreover, $\mathbf{\hat{b}}_u[k]^H$ is the $u^{th}$ row of the ZF combiner $\mathbf{\hat{B}}[k]$. Furthermore, $\mathbf{{\tilde{h}}}_u[k]$ and $\mathbf{{\hat{h}}}_u[k]$ are equal to the $u^{th}$ column of the matrices $\mathbf{\tilde{H}}[k]$ and $\mathbf{\hat{H}}[k]$, respectively. In addition, $\mathbf{\tilde{e}}_u[k]$ corresponds to the $u^{th}$ column of the channel error matrix $\mathbf{\tilde{H}}[k]-\mathbf{\hat{H}}[k]$.

%A lower bound on the ergodic capacity per user, for which the receiver has access to the side information $\mathbf{\hat{H}}\triangleq{\mathbf\{\mathbf{\hat{H}}}[0],\mathbf{\hat{H}}[1], \ldots, \mathbf{\hat{H}}[N-1]\}$, is derived in the following proposition.

A lower bound on the ergodic capacity per user, for which the receiver has access to the side information $\mathbf{\hat{H}}\triangleq{\mathbf\{\mathbf{\hat{H}}}[0],\mathbf{\hat{H}}[1], \ldots, \mathbf{\hat{H}}[N-1]\}$, an SINDR expression for the data symbol of $u^{th}$ user at the $k^{th}$ subcarrier, namely $\gamma_u[k]$, which is defined in \cite[eq. (2.46)]{Marzetta1088544}, is found using (\ref{eqn:rec_sig_compnts})-(\ref{eqn:rec_sig_compnts3}) in the following proposition.

\begin{prop} \label{prop:tight_SINDR}
A lower bound on the ergodic capacity per user, treating $\mathbf{\hat{H}}$ as side information, which is denoted by $R$, can be calculated as
\begin{equation}
\label{eqn:erg_ach_rate}
R=\dfrac{1}{U|\mathcal{K}_D|}\mathbb{E}_{\hat{\mathbf{H}}}\left\lbrace\sum_{u=1}^{U}\sum_{k\in\mathcal{K}_D}\mathrm{log}_2\left(1+\gamma_u[k]\right)\right\rbrace,
\end{equation}
where
\begin{align}
\gamma_u[k]&\triangleq\dfrac{|\mathbb{E}[g_u[k]'|\mathbf{\hat{H}}]|^2}{\text{Var}\left[g_u[k]'|\mathbf{\hat{H}}\right]+\text{Var}\left[i_u[k]+n_u[k]+q_u[k]|\mathbf{\hat{H}}\right]}=\dfrac{\rho_d^2N|\mathbf{\hat{b}}_u[k]^H\mathbf{A}{\mathbf{\hat{h}}_u}[k]|^2}{I_u[k]+N_u[k]+Q_u[k]}, \label{eqn:sinr_cond_h_est}
\end{align}
\begin{align}
I_u[k]&=\rho_d^2N\sum_{z\neq u,z\in\mathcal{U}_d}\left[|\mathbf{{\hat{b}}}_u[k]^H\mathbf{A}{\mathbf{\hat{h}}_z}[k]|^2+U\sigma_e^2\rho_d^2||\mathbf{{\hat{b}}}_u[k]^H\mathbf{A}||^2\right],\label{eqn:sinr_est_inside_terms}\\
N_u[k]&=NN_o||\mathbf{{\hat{b}}}_u[k]^H\mathbf{A}||^2, Q_u[k]=\mathbf{{\hat{b}}}_u[k]^H\mathbf{C}_{\mathbf{q}[k]}\mathbf{{\hat{b}}}_u[k],
\end{align}
when $\mathbf{\hat{H}}$ is the LMMSE channel estimate. In (\ref{eqn:sinr_est_inside_terms}), $\sigma_e^2$ is the LMMSE channel estimation error variance, which is as defined in Section \ref{sec:chan_est}.
\begin{proof}
See Appendix \ref{apdx:prop3_proof}.
\end{proof}
\end{prop}

%%\begin{align}
%%\gamma_u[k]&=\dfrac{|\mathbb{E}\left[\hat{x}_u[k]|\underline{\mathbf{H}}\right]|^2}{\text{Var}\left[\hat{x}_u[k]|\underline{\mathbf{H}}\right]}\\
%%\label{eqn:SINDR}&=\dfrac{\rho_d^2|\mathbf{\tilde{b}}_u[k]^H\mathbf{A}{\mathbf{\tilde{h}}_u}[k]|^2}{I_u[k]+N_u[k]+Q_u[k]}
%%\end{align}
%%where $\underline{\mathbf{H}}$ is the conditional variance of $\hat{x}_u[k]$ conditioned on $\mathbf{H}$. 
%The corresponding lower bound on the ergodic capacity per user, treating $\mathbf{\hat{H}}$ as side information, which is denoted by $R$, can also be found as \cite[eq. (2.46)]{Marzetta1088544}
%\begin{equation}
%\label{eqn:erg_ach_rate}
%R=\dfrac{1}{U|\mathcal{K}_D|}\mathbb{E}_{\hat{\mathbf{H}}}\left\lbrace\sum_{u=1}^{U}\sum_{k\in\mathcal{K}_D}\text{log}_2\left(1+\gamma_u[k]\right)\right\rbrace.
%\end{equation}

Moreover, to find the bit error rate (BER) for gray coded $K$-ary phase shift keying modulation ($K$-PSK), we approximate $\mathbf{\hat{x}}[k]$ as a complex Gaussian random variable, so that \cite{Goldsmith_WC}
\begin{align}
\mathrm{BER}\approx&\dfrac{1}{U|\mathcal{K}_D|}\mathbb{E}_{\mathbf{\hat{H}}}\left\lbrace\sum_{u=1}^{U}\sum_{k\in\mathcal{K}_D}\frac{2}{\mathrm{log}_2(K)}\times \left(1-\Phi\left(\sqrt{\gamma_u[k]\mathrm{log}_2(K)}\mathrm{sin}\left(\frac{\pi}{K}\right)\right)\right)\right\rbrace.\label{eqn:BER_exp}
\end{align}
%An ergodic capacity expression could also be found by using the SINDR expression in (\ref{eqn:SINDR}) directly. However, for the imperfect CSI case, it will be shown that the signal part in the SINDR expression in (\ref{eqn:SINDR}), whose average power is found as in the nominator of (\ref{eqn:BER_exp}), is correlated with the interference and noise terms in (\ref{eqn:SINDR}). The SINDR expression in (\ref{eqn:SINDR}) is useful to calculate the error-rate performance of the ZF receiver characterized by (\ref{eqn:ZF_detection}), which does not take into account the correlation between the channel estimates and the noise and interference components in the received signal. However, for the SINDR expression of a more optimal receiver taking into account the mentioned correlation, the signal part and the total effective noise (multi-user interference+ quantization noise + thermal noise) parts should be uncorrelated.

\color{black} The BER calculation formula in (\ref{eqn:BER_exp}) is applicable for a multi-band interference environment, as the calculation of the SINDR $\gamma_u[k]$ involves the received power from any interfering band. \color{black} As detailed in Footnote~\ref{ftnt:Quant_inp_gauss}, the approximation error in assuming Gaussian inputs for the quantizer is very limited, thus $R$ and the BER expressions in (\ref{eqn:erg_ach_rate}) and (\ref{eqn:BER_exp}) calculated using $\gamma_u[k]$ in (\ref{eqn:sinr_cond_h_est}) are expected to approximate the simulation based results precisely ($\gamma_u[k]$ is the precise SINDR expression mentioned in Section~\ref{sec:intro}). \color{black}Although (\ref{eqn:erg_ach_rate}) and (\ref{eqn:BER_exp}) are useful to calculate values of the ergodic capacity and BER, they are not able to provide clear insights into the system performance and system parameters such as $M$, $N$ or $\rho_d^2/\rho_i^2$. Therefore, we propose a more tractable approximation of $\gamma_u[k]$ in Proposition \ref{prop:approx_SINDR}, in which the conditioning on $\mathbf{\hat{H}}$ will be dropped. From an information theoretic view, this will correspond to the case that the channel estimates are used to perform ZF combining by a first party, but the channel estimate knowledge is not conveyed to a second party, which performs error correction decoding on the ZF output without the knowledge of channel estimates \cite{Marzetta1088544}. The corresponding use-and-then-forget ergodic capacity bound, namely $R'$, is found in the following proposition.

\begin{prop} \label{prop:approx_SINDR} 
Use-and-then-forget ergodic capacity bound $R'$ can be computed as
\begin{align}
R'=\dfrac{1}{U|\mathcal{K}_D|}\sum_{u=1}^{U}\sum_{k\in\mathcal{K}_D}&\text{log}_2\left(1+\gamma_u'[k]\right),
\label{eqn:approx_ach_rate_prop_4}\\
\gamma_u'[k]=\dfrac{|\mathbb{E}[g_u[k]']|^2}{\text{Var}[g_u[k]']+\text{Var}[i_u[k]+n_u[k]+q_u[k]]}&\approx \dfrac{\rho_d^2(1-\sigma_e^2)(M-U)G^2}{\left(U\sigma_e^2\rho_d^2G^2+2-4/\pi+N_oG^2\right)},\label{eqn:approx_SINDR_expression}
\end{align}
in which
$G=2/\sqrt{\pi}\left(\left(|\mathcal{K}_D|U\rho_d^2+|\mathcal{K}_I|I\rho_i^2\right)/N+N_o\right)^{-0.5}$. The approximation error goes to zero as $L$ gets larger and $|\mathcal{K}_D|+|\mathcal{K}_I|$ approach to $N$ (as oversampling rates gets lower) when $\rho_d^2\approx\rho_i^2$.
\begin{proof}
See Appendix \ref{apdx:prop4_proof}.
\end{proof}
\end{prop}
\color{black} Note that there is no need for any Monte-Carlo based simulation to obtain $\gamma_u'[k]$, since it can be calculated by just plugging the system parameters into the rightmost expression in (\ref{eqn:approx_SINDR_expression}). $\gamma_u'[k]$ can be used in place of $\gamma_u[k]$ in (\ref{eqn:BER_exp}) or in (\ref{eqn:approx_ach_rate_prop_4}) to calculate BER and achievable rate, again without any Monte-Carlo simulations (as averaging over $\mathbf{\hat{H}}$ is already performed by analysis to obtain $\gamma_u'[k]$). \color{black} Owing to the simple form of $\gamma_u'[k]$ in (\ref{eqn:approx_SINDR_expression}), insights into the system performance and parameters can be obtained as follows. \color{black} From Proposition \ref{prop:approx_SINDR}, it is obvious that it is always possible to increase the SINDR by increasing the number of antennas $M$. The impact of the oversampling rate can be deduced by considering the case $|\mathcal{K}_D|$ and $|\mathcal{K}_I|$ are fixed as the block length $N$ increases which by definition is equivalent to an increase in the oversampling rates $\beta_I$ and $\beta_D$. It can be shown that $\gamma_u'[k]$ is increasing with $N$ by considering $\gamma_u'[k]=\rho_d^2(M-U)(1-\sigma_e^2)\overline{\gamma}_u[k]$, where  $\overline{\gamma}_u[k]=G^2/\left(2-4/\pi+\left(N_o+U\sigma_e^2\rho_d^2\right)G^2\right)$. Since $G$ and $G^2$ are increasing with $N$, it follows that $\overline{\gamma}_u[k]=G^2/\left(2-4/\pi+\left(N_o+U\sigma_e^2\rho_d^2\right)G^2\right)$ also increases with $N$, which in turn means that $\gamma_u'[k]$ increases with $N$. The channel estimation error will be calculated in (\ref{eqn:analyt_est_err}) in the next section, which can similarly be shown to decrease with the oversampling rates.

\subsection{Channel Estimation}
\label{sec:chan_est}
In this section, the details of channel estimation under quantization will be presented. There are many channel estimation techniques for massive MIMO systems with low-resolution ADCs \cite{Mollen_wideband,nML_MM_MIMO,jacobsson_recent,mixed_adc,larsson_MM_1bitADC}.  In none of those studies, a channel estimation scheme under ACI is discussed. In this study, we will propose an LMMSE channel estimation based on Bussgang decomposition. The estimation technique is the extension of the channel estimation technique in \cite{Mollen_wideband}. For the channel estimation phase, orthogonal pilot sequences are transmitted at some of the subcarriers involved. The set of subcarriers for which pilot signals are transmitted is denoted by $\mathcal{K}_P$. Moreover, it will be assumed that the interferers at the adjacent channel band will be transmitting data symbols from a finite cardinality set through the subcarriers in $\mathcal{K}_I$ during the channel estimation phase of the users in the desired band. The transmitted time domain pilot signal of the $u^{th}$ user of length $N_p$, where $u\in\mathcal{U}_D$, can be expressed as follows:
\begin{equation}
p_u[n]=\begin{cases} \dfrac{\rho_p}{\sqrt{N_p}}\sum_{k\in\mathcal{K}_P}\tilde{\theta}_u[k]e^{j2\pi nk/N_p}, & \text{if } u\in\mathcal{U}_D, \\
\dfrac{\rho_i}{\sqrt{N_p}}\sum_{k\in\mathcal{K}_I}\tilde{s}_u[k]e^{j2\pi nk/N_p}, & \text{if } u\in\mathcal{U}_I,
  \end{cases}
\end{equation}
\begin{equation}
\tilde{\theta}_u[k]=
\begin{cases}
0, & \text{if } (k \text{ mod } U)+1\neq u,\\
\sqrt{U}e^{j\phi_u[k]}, & \text{if } (k \text{ mod } U)+1=u.
\end{cases}
\label{eqn:pilot_phase}
\end{equation}
Here, the phases $e^{j\phi_u[k]}$ are known by the base station. The selection of these phases affect the estimation performance. They are selected from the uniform distribution as suggested in \cite{Mollen_wideband} since when they are selected as constant, that is, when $e^{j\phi_u[k]}=C\ \forall u,k$, the transmitted signal by $u^{th}$ user $p_u[n]\neq0$ only when $n=\mu N_p/, \mu\in \mathbb{Z}$. Otherwise, $p_u[n]=0$. This means that users do not transmit anything most of the time, which limits the average transmit power due to the peak power limitation of the power amplifiers in the transmitter side. Introduction of non-constant phases, one example of which is when they are selected from uniform distribution ($0,2\pi$), avoids this problem. Since the channel convolution matrix is circulant due to CP, the received signal at the $m^{th}$ antenna and the $k^{th}$ subcarrier can be expressed as follows:
\begin{align}
\tilde{y}_m[k]=
\begin{cases}\rho_p\sqrt{N_pU}\sum_{u\in\mathcal{U}_D}\tilde{h}_{m,u}[k]\tilde{\theta}_u[k]+z_m[k], & \text{if } k\in\mathcal{K}_P,\\
\rho_i\sqrt{N_pU}\sum_{u\in\mathcal{U}_I}\tilde{h}_{m,u}[k]\tilde{s}_u[k]+z_m[k], & \text{if } k\in\mathcal{K}_I,
\label{eqn:unquant_rec_sig_pilot}
\end{cases}
\end{align}
where $\tilde{s}_u[k]$'s for $k\in\mathcal{K}_I$, $u\in\mathcal{U}_I$ represent the random data symbols transmitted by the interfering band users, $\tilde{h}_{m,u}[k]$ is the element of matrix $\mathbf{\tilde{H}}[k]$ at its $m^{th}$ row and $u^{th}$ column, and $z_m[n]$ represents the additive white noise term at the $m^{th}$ receive antenna of spectral density $N_pN_o$. Due to (\ref{eqn:pilot_phase}), it can be written that
\begin{equation}
\label{eqn:rec_sig_pilot}
\tilde{y}_m[k]=
\begin{cases}\rho_p\sqrt{N_pU}\tilde{h}_{m,f(k)}[k]e^{j\phi_{f(k)}[k]}+z[k], & \text{if } k\in\mathcal{K}_P,\\
\rho_i\sqrt{N_pU}\sum_{u\in\mathcal{U}_I}\tilde{h}_{m,u}[k]\tilde{s}_u[k]+z[k], & \text{if } k\in\mathcal{K}_I,
\end{cases}
\end{equation}
where $f(k)=(k \text{ mod } U)+1$. Defining the quantized observation vector $v_m[n]\triangleq\mathrm{sign}(\Re\{{y}_m[n]\})+j\mathrm{sign}(\Im\{{y}_m[n]\})$ in time domain, the quantized observation $\tilde{v}_m[k]$ in the frequency domain can be expressed as
\begin{equation}
\tilde{v}_m[k]=\sum_{k=0}^{N_p-1}v_m[n]e^{j2\pi nk/N_p}.
\end{equation}
Here $y_m[n]$ is the IDFT of $\tilde{y}_m[k]$. Defining $\mathbf{v}[n]\triangleq[v_1[n]\ v_2[n] \ \cdots \ v_M[n]\ ]^T$ and $\mathbf{y}[n]\triangleq[\ y_1[n] \ y_2[n] \ \cdots \ y_M[n] \ ]^T$, it can be written that
\begin{equation}
\mathbf{v}[n]=\mathbf{A}'\mathbf{y}[n]+\mathbf{q}'[n],
\label{eqn:bussg_decom_pilot}
\end{equation}
where the selection $\mathbf{A}'=\sqrt{4/\pi}\mathrm{diag}\left(\mathbf{C}_y[0]\right)^{-0.5}$ makes the quantization noise $\mathbf{q}'[n]$ to be uncorrelated with the unquantized observation vector $\mathbf{y}[n]$\footnote{Here, the quantizer inputs are again assumed to be Gaussian due to the same reasoning discussed in Footnote~\ref{ftnt:Quant_inp_gauss}.}. Let $\mathbf{\underline{y}}\triangleq[\mathbf{y}[N-1]^T \  \mathbf{y}[N-2]^T\  \ldots \ \mathbf{y}[0]^T]^T$. For the simplicity of the channel estimation part, $\mathbf{C}_{\underline{y}}$ is approximated as
\begin{align}
\label{eqn:avg_rec_pow}
\mathbf{C}_{\underline{y}}&=\underline{\mathbf{H}}\mathbf{R_{\underline{p}}}\underline{\mathbf{H}}^H+N_o\mathbf{I}\approx\left(1/N_p(\rho_p^2|\mathcal{K}_p|U+\rho_i^2|\mathcal{K}_I|U)+N_o\right)\mathbf{I},
\end{align}
where $\mathbf{R_{\underline{p}}}=\mathbb{E}[\mathbf{\underline{p}}\mathbf{\underline{p}}^H]$, in which, $\mathbf{\underline{p}}=[\mathbf{p}[N-1]^T \  \mathbf{p}[N-2]^T\  \ldots \ \mathbf{p}[0]^T]^T$, where $\mathbf{p}[n]$ is an $M\times1$ column vector whose $u^{th}$ element is $p_u[n]$. Without the approximation, $MN\times MN$ matrix $\mathbf{C}_{\underline{y}}$ will be non-diagonal in general, which implies that $MN\times MN$ quantization noise covariance matrix will be non-diagonal. This will require taking the inverse of such a large matrix for LMMSE channel estimation as in \cite{Li_perf_analysis} for frequency selective channel, which is computationally exhaustive. However, the approximation error goes to zero as $L$ grows large which can be shown similarly as performed for $\mathbf{C_r}[0]$ in Appendix~\ref{apdx:prop4_proof}. Such an approximation is also adopted in \cite{Mollen_wideband}. Taking the DFT of the quantized observation vector $\mathbf{v}[n]$, it is found that
\begin{equation}
\mathbf{\tilde{v}}[k]=G'\mathbf{\tilde{y}}[k]+\mathbf{\tilde{q}}'[k],
\label{eqn:bussg_decom_pilot_freq_dom}
\end{equation}
where $G'=2/\sqrt{\pi}\left((\rho_p^2|\mathcal{K}_p|U+\rho_i^2|\mathcal{K}_I|I)/N_p+N_o\right)^{-0.5}$. (\ref{eqn:bussg_decom_pilot_freq_dom}) along with (\ref{eqn:rec_sig_pilot}) implies that
\begin{align}
\label{eqn:rec_sig_pilot_quantizd}
\tilde{y}_m[k]=\rho_pG'\sqrt{N_pU}h_{m,f(k)}[k]e^{j\phi_{f(k)}[k]}+G'z_m[k]+p_m[k],
\end{align}
where $k\in\mathcal{K}_P$ and $p_m[k]$ is the $m^{th}$ element of the DFT of $\mathbf{p}[n]$. It can be seen from (\ref{eqn:rec_sig_pilot_quantizd}) that the observation $\tilde{y}_m[vU+u-1]$, when the noise terms are omitted, is a phase rotated and scaled version of the channel coefficient $\tilde{h}_{m,u}[vU+u-1]$ for user $u$, sampled with a sampling period of $U$, as $f(vU+u-1)=(vU+u-1 \text{ mod } U)+1=u$. These samples will be denoted by $\check{h}_{m,u}[v]\triangleq \tilde{h}_{m,u}[vU+u-1]$. According to the Nyquist sampling theorem, if $N_p$ satisfies
\begin{equation}
N_p\geq UL,
\end{equation}
it is possible to obtain the channel coefficients without any aliasing. Again for the simplicity of the channel estimation part, the covariance matrix of $\mathbf{q}'[n]$ will be approximated as a diagonal matrix $(2-4/\pi)\mathbf{I}$, with approximation error going to zero as $L$ grows large for low oversampling rates and $\rho_p^2\approx\rho_i^2$ as discussed in Appendix~\ref{apdx:prop4_proof}. Under this approximation, the LMMSE estimate for the channel coefficient $\tilde{h}_{m,u}[vU+u-1]$, namely $\tilde{h}_{m,u}[vU+u-1]^*$, is found as
\begin{align}
\tilde{h}_{m,u}[vU+u-1]^*=\dfrac{e^{-j\phi_{u}[vU+u-1]}\tilde{y}_m[vU+u-1]}{\rho_pG'\sqrt{N_pU}\left(1+N_o/(\rho_p^2U)+P/\left(\rho_p^2U\left(G'\right)^2\right)\right)},\label{eqn:LMMSE_est}
\end{align}
where the quantization distortion variance $P=\mathbb{E}[|p_m[k]|^2]\approx2-4/\pi$. Here, the channel estimation error $\sigma_e^2$ can also be found as
\begin{equation}
\sigma_e^2=1-\frac{1}{\rho_p\sqrt{N_pU}\left(1+N_o/(\rho_d^2U)+P/\left(\rho_d^2U\left(G'\right)^2\right)\right)}.
\label{eqn:analyt_est_err}
\end{equation}

The parameter $\sigma_e^2$ will be used in the performance analysis of the investigated uplink system model in this paper for imperfect CSI and the results from the analysis will be compared to the simulated results. As can be noted in (\ref{eqn:LMMSE_est}), for the $u^{th}$ user, we only have the channel coefficients estimates for $\check{h}_{m,u}[v]\triangleq h_{m,u}[vU+u-1]^*$, sampled with a period of $U$. To obtain the remaining channel coefficients, $\check{h}_{m,u}[v]$ can be upsampled by $U$. There are many possibilities to upsample $\check{h}_{m,u}[v]$. The one adopted in this study is the spline interpolation \cite{de1978practical}.

\color{black} The most important parameter through which the ACI is taken into account in the proposed channel estimation method is through the factor $G'$ in (\ref{eqn:LMMSE_est}). By definition, $G'$ decreases with increasing total ACI power $\rho_i^2|\mathcal{K}_I|I/N_p$. The distortion caused by the quantization can be regarded to have two components, the additive quantization noise distortion $\mathbf{\tilde{q}}'[k]$ and the magnitude distortion $G'$ in (\ref{eqn:bussg_decom_pilot_freq_dom}). Under the aforementioned approximations, the power of the additive quantization noise $\mathbf{\tilde{q}}'[k]$ does not change with the ACI power. However, as $G'$ decreases with increasing ACI power, the power of the signal part $G'\mathbf{\tilde{y}}[k]$ in (\ref{eqn:bussg_decom_pilot_freq_dom}) diminishes, resulting in a reduced signal power compared to the quantization noise power. Therefore, a worse estimation error performance can be expected. This can also be interpreted from (\ref{eqn:analyt_est_err}). The channel estimation error variance $\sigma_e^2$ in (\ref{eqn:analyt_est_err}) increases as $G'$ decreases with increasing ACI power. Regarding how the estimator combats with the degredation due to ACI can be inferred from (\ref{eqn:LMMSE_est}). Neglecting the $P/(\rho_p^2U(G')^2)$ term in the denominator in (\ref{eqn:LMMSE_est}), it can be stated that as the ACI power is increased, which in turn decreases $G'$ and reduces the signal component $G'\mathbf{\tilde{y}}[k]$ in (\ref{eqn:bussg_decom_pilot_freq_dom}), the estimator tries to cancel this effect by multiplying the observation by $1/G'$ (note the $G'$ factor in the denominator in (\ref{eqn:LMMSE_est})). However, such a normalization (multiplication by $1/G'$ when $G'$ is smaller than $1$) results in the enhancement of the quantization noise $\mathbf{\tilde{q}}'[k]$, thus the cancellation of the magnitude distortion $G'$ should be balanced with the quantization noise enhancement. This balancing is performed through the $P/\rho_p^2U(G')^2$ factor in the estimator in (\ref{eqn:LMMSE_est}). \color{black}
\section{ADCs with higher than one-bit resolution}
\label{sec:higher_res_ADC}
In this section, the details for the performance analysis for quantizers with more than one-bit resolution is presented. To begin with, we define the set of quantizer output values $\mathcal{L}=\{\ell_0,\ell_1,\ldots,\ell_{L'-1}\}$, where $L'=2^q$ is the number of possible quantizer output values $q$ being the number of ADC bits. Moreover, the quantization thresholds can also be characterized by the set $\mathcal{B}=\{b_0, b_1, \ldots, b_{L'}\}$, where $-\infty =b_0 < b_1 < \cdots < b_{L'}=\infty$. The quantization function $\mathcal{Q}(.)$ is a point in the function space $\mathbb{C}^M\rightarrow {\Upsilon}^M$, where $\mathbb{C}^M$ denotes the complex vector space of dimension $M$ and $\Upsilon=\mathcal{L}\times\mathcal{L}$ is the set of possible quantizer output values (the cartesian product $\mathcal{L}\times\mathcal{L}$ represents the combination of the outputs of the pair of ADCs quantizing the real and imaginary parts of the received signals separately). The $i^{th}$ element of the quantizer output, namely $\mathcal{Q}(\mathbf{x})_i$, where $\mathbf{x}$ is the quantizer input vector of size $M\times1$ can be expressed as
\begin{equation}
\mathcal{Q}(\mathbf{x})_i=\left(\ell_{f'\left(\Re\left(x_i\right)\right)}, \ell_{f'\left(\Im\left(x_i\right)\right)}\right),
\end{equation}
where $f'(\Re(x))=k\in \{0,1,\ldots,L'-1\}$ which satisfies $b_k\leq \Re(x) < b_{k+1}$. Similarly $f(\Im(x))=m\in\{0,1,\ldots,L'-1\}$ which satisfies $b_m\leq \Im(x) < b_{m+1}$. As an example, the possible quantizer output values $\ell_i=\Delta(i-L'/2+1/2)$, $i=0, 1, \ldots, L'-1$, whereas the quantization thresholds ($b_i$) can be found as $b_i=\Delta\left(i-\frac{L'}{2}\right)$, $i=1,2,\ldots,L'-1$ for a uniform midrise quantizer ($b_0=-\infty, b_{L'}=\infty$ as previously specified).

An important point in the design of the quantizer is the selection of the step size $\Delta$. In fact, AGC will dynamically adjust the gain of the input signal to ADC according to the received signal power in order that it fits the input signal range of the ADC. This will correspond to the approach in this study in which the step size is selected according to the received signal power levels, which is assumed to stay nearly the same over a coherence interval. This will result in a fixed step size during a coherence interval, enabling a tractable analysis.

There are two main considerations in the design of the step size $\Delta$. If the step size is selected to be small for the average received signal power level, the probability that the input signal is clipped will be high and cause a distortion, referred to as \textit{overload} distortion. On the other hand, if a large step size is preferred to avoid clipping or overload distortion, this will result in a granular distortion, causing a large range of input signal level to be mapped to the same level. Therefore, step size should be selected properly to balance the aforementioned granular and overload distortions. The amount of the two distortions will affect the validity of the assumptions in the performance analysis, as will be discussed in the subsequent parts of this section.

To begin with the analysis, matrix $\mathbf{A}$ in (\ref{eqn:no_time_index_quant_sig}) should be evaluated. According to Bussgang decomposition, $\mathbf{A}=\mathbf{C_{d[n]r[n]}C_{r[n]}^{-1}}$, where $\mathbf{C_{d[n]r[n]}}=\mathbb{E}[\mathbf{d[n]r[n]^H]}$ and $\mathbf{C_{r[n]}}=\mathbb{E}[\mathbf{r[n]r[n]^H}]$. For the example case of midrise uniform quantizer with Gaussian inputs\footnote{\label{ftnt:mult_bit_gauss_fn}The multi-bit quantizer input is also assumed to be Gaussian, which is accurate owing to the same reasoning discussed in Footnote \ref{ftnt:Quant_inp_gauss}.} \cite{jacobsson_aggragate},
\begin{align}
\mathbf{A}=\frac{\Delta}{\sqrt{\pi}}\mathrm{diag}\left(\mathbf{C}_r[0]\right)^{-0.5}\times \sum_{i=1}^{2^{q}-1}\mathrm{exp}\left(-{\Delta}^2\left(i-2^{q-1}\right)^2\mathrm{diag}\left(\mathbf{C}_r[0]\right)^{-0.5}\right).\label{eqn:mtx_A_for_multi_bit}
\end{align}
As $\mathbf{C}_r[0]$ in (\ref{eqn:mtx_A_for_multi_bit}) can be found using Proposition~\ref{prop:time_dom_correlation}, matrix $\mathbf{A}$ can be calculated using (\ref{eqn:mtx_A_for_multi_bit}) for multi-bit quantizer case. What remains is the calculation of the covariance matrix of the quantization noise $\mathbf{C}_{\mathbf{q}[k]}$. The difficulty with the calculation of this matrix stems from the fact that there is no closed form expression for the relation between the quantizer input and output covariance matrices for multi-bit quantizers as for the one-bit quantizer in (\ref{eqn:arcsine_rule}), which was referred to as the arcsine law. However, a diagonal approximation can be made, for which all non diagonal entries of the covariance matrix $\mathbf{C_d}[0]$ are assumed to be zero and the $m^{th}$ diagonal entry of $\mathbf{C_d}[0]$, namely $\mathbb{E}\left[|d_m[n]|^2\right]$, can be found as follows:
\begin{align}
\mathbb{E}\left[|d_m[n]|^2\right]=2\sum_{i=0}^{L'-1}\ell_i^2\mathbb{P}\left[b_i\leq \Re(r_m) < b_{i+1}\right]=2\sum_{i=0}^{L'-1}\ell_i^2\left(\Phi\left({\sqrt{2}b_{i+1}}/{{\sigma_{r_m}}}\right)-\Phi\left({\sqrt{2}b_i}/{\sigma_{r_m}}\right)\right),\label{eqn:quant_inp_gauss_assump}
\end{align}
where $\sigma_{r_m}^2$ is the $m^{th}$ diagonal element of $\mathrm{diag}(\mathbf{C_r}[0])_{m,m}$ corresponding to the quantizer input variance at the $m^{th}$ antenna. In (\ref{eqn:quant_inp_gauss_assump}), it is assumed that the quantizer input has a Gaussian distribution\footnotemark[\getrefnumber{ftnt:mult_bit_gauss_fn}]. We will denote the diagonal matrix whose diagonal entries are equal to the diagonal entries of $\mathbf{C_d}[0]$ as $\mathbf{C}^{\mathrm{diag}}_\mathbf{d}[0]$. After finding $\mathbf{C}^{\mathrm{diag}}_\mathbf{d}[0]$ from (\ref{eqn:quant_inp_gauss_assump}), the diagonal approximation for the $\mathbf{C_q}[0]$, namely $\mathbf{C_q^{{\mathrm{diag}}}}[0]$, can be found from (\ref{eqn:dist_simplified}) as
\begin{equation}
\label{eqn:quant_distortion_cov}
\mathbf{C_q^{{\mathrm{diag}}}}[0]=\mathbf{C}^{\mathrm{diag}}_\mathbf{d}[0]-\mathbf{A}\mathrm{diag}(\mathbf{C_r}[0])\mathbf{A}^H.
\end{equation}
The covariance matrices of quantization noise for nonzero lags, namely $\mathbf{C_q^{{\mathrm{diag}}}}[m]$, $m\neq0$, are also assumed to be zero for multi-bit quantizers, which means that the correlation in time for the quantization noise is assumed to be zero. This assumption fails to be valid for very low ADC resolutions \cite{jacobsson_lin_prec} \color{black} or for high oversampling rates, as discussed in Appendix~\ref{apdx:prop4_proof}, \color{black}  yet, it provides more accurate results as the number of quantization bits is increased \cite{jacobsson_lin_prec}, when clipping or overload distortion occurs with low probability. To ensure this, we will choose the step size $\Delta$ small enough as will be discussed shortly. After finding $\mathbf{C_q^{{\mathrm{diag}}}}[0]$, $\mathbf{C}_q[k]$ for the diagonal approximation case, which is referred to as $\mathbf{C}^{\mathrm{diag}}_q[k]$, can be found using Proposition~\ref{prop:error_cov_freq} as follows:
\begin{align}
\mathbf{C}^{\mathrm{diag}}_q[k]=\mathrm{DFT}_{N,k}\left\lbrace\mathbf{\Gamma}[m]\right\rbrace+\mathrm{DFT}_{N,k}\left\lbrace\mathbf{\Gamma}[m]\right\rbrace^*-N\mathbf{C_{\tilde{q}}}[0]=N\mathbf{C_{\tilde{q}}}[0],
\label{eqn:error_cov_diag}
\end{align}
as it is assumed for multi-bit quantizers that $\mathbf{\Gamma}[m]=(N-m)\mathbf{C_q}[m]\approx(N-m)\mathbf{C_q^{{\mathrm{diag}}}}[m]=0$ for $m\neq0$. Then, $\mathbf{C_q^{{\mathrm{diag}}}}[k]$ can be used to find the SINDR expression in Proposition~\ref{prop:tight_SINDR}, which can be employed to find the error-rate performance using (\ref{eqn:BER_exp}). To ensure that the overload distortion is negligible, we adjust the step size $\Delta$ as follows:
\begin{equation}
\label{eqn:step_size_adjs_no_clip}
\Delta=2A_{max}/L',
\end{equation}
where the maximum quantizer output level $A_{max}$ is adjusted as $A_{max}=\sqrt{G/2}\left(1-\Phi\left(P_{c}/2\right)\right)$, where $G$ is as defined in Proposition~\ref{prop:approx_SINDR}, which corresponds to the average received power and $P_{c}$ is the desired probability that a clipping occurs. Obviously, the received signal is also assumed to be Gaussian distributed in the adjustment of the step size, which is an accurate assumption according to Footnote~\ref{ftnt:Quant_inp_gauss}. The clipping probability will be chosen as a small number, owing to its impact on the validity of the diagonal approximations involved in the analysis. Imperfect CSI case for multi-bit quantizer cannot be covered due to limited space and is left for future work.

\color{black}To see the effect of oversampling, number of ADC bits and number of antennas on the SINDR for the multi-bit quantizer case, the parameter $G$ in (\ref{eqn:approx_SINDR_expression}) can be found using (\ref{eqn:mtx_A_for_multi_bit}) as 
\begin{equation}
\label{eqn:G_multi-bit}
G=\frac{\Delta}{\sqrt{\pi}}\left(\lambda\right)^{-0.5}\sum_{i=1}^{2^{q}-1}\mathrm{exp}\left(-{\Delta}^2\left(i-2^{q-1}\right)^2\left(\lambda\right)^{-0.5}\right),
\end{equation}
where $\lambda=\left(|\mathcal{K}_D|U\rho_d^2+|\mathcal{K}_I|I\rho_i^2\right)/N+N_o$. The SINDR expression in (\ref{eqn:approx_SINDR_expression}) will be the same for the multi-bit quantizer case except that the parameter $G$ in (\ref{eqn:approx_SINDR_expression}) is found according to (\ref{eqn:G_multi-bit}) and the quantization noise variance $\text{Var}[q_u[k]]$ will also be another constant less than $2-4/\pi$, which is decreasing with the number of ADC resolution bits, but will not change with the number of antennas or the oversampling rate. As mentioned before, increasing $N$ while $|\mathcal{K}_D|$ and $|\mathcal{K}_I|$ are fixed corresponds to an increase in the oversampling rates. In such case, $G$ is increased as $\lambda$ decreases with $N$. Therefore, the same discussion that SINDR $\gamma_u'[k]$ increases with $G$ or the oversampling rates for one-bit ADC also applies for the multi-bit quantizer. Moreover, due to the $(M-U)$ factor in (\ref{eqn:approx_SINDR_expression}), it is also possible to increase SINDR by increasing the number of antennas $M$. Furthermore, since the step size $\Delta$ decreases when the number of ADC bits is increased, this corresponds to an increase in $G$ according to (\ref{eqn:G_multi-bit}), and a decrease in the quantization noise variance $\text{Var}[q_u[k]]$, which in turn results in an increase in the SINDR $\gamma_u'[k]$ in (\ref{eqn:approx_SINDR_expression}). Therefore, it can be stated that SINDR will increase when the oversampling rates, number of antennas and ADC bits are increased for the multi-bit quantizer case, in line with the intuition. \color{black}

\section{Simulation Results}
\label{sec:sim_res}
\begin{figure*}
\centering
\begin{subfigure}[t]{0.5\textwidth}
\centering
\includegraphics[width=0.95\textwidth]{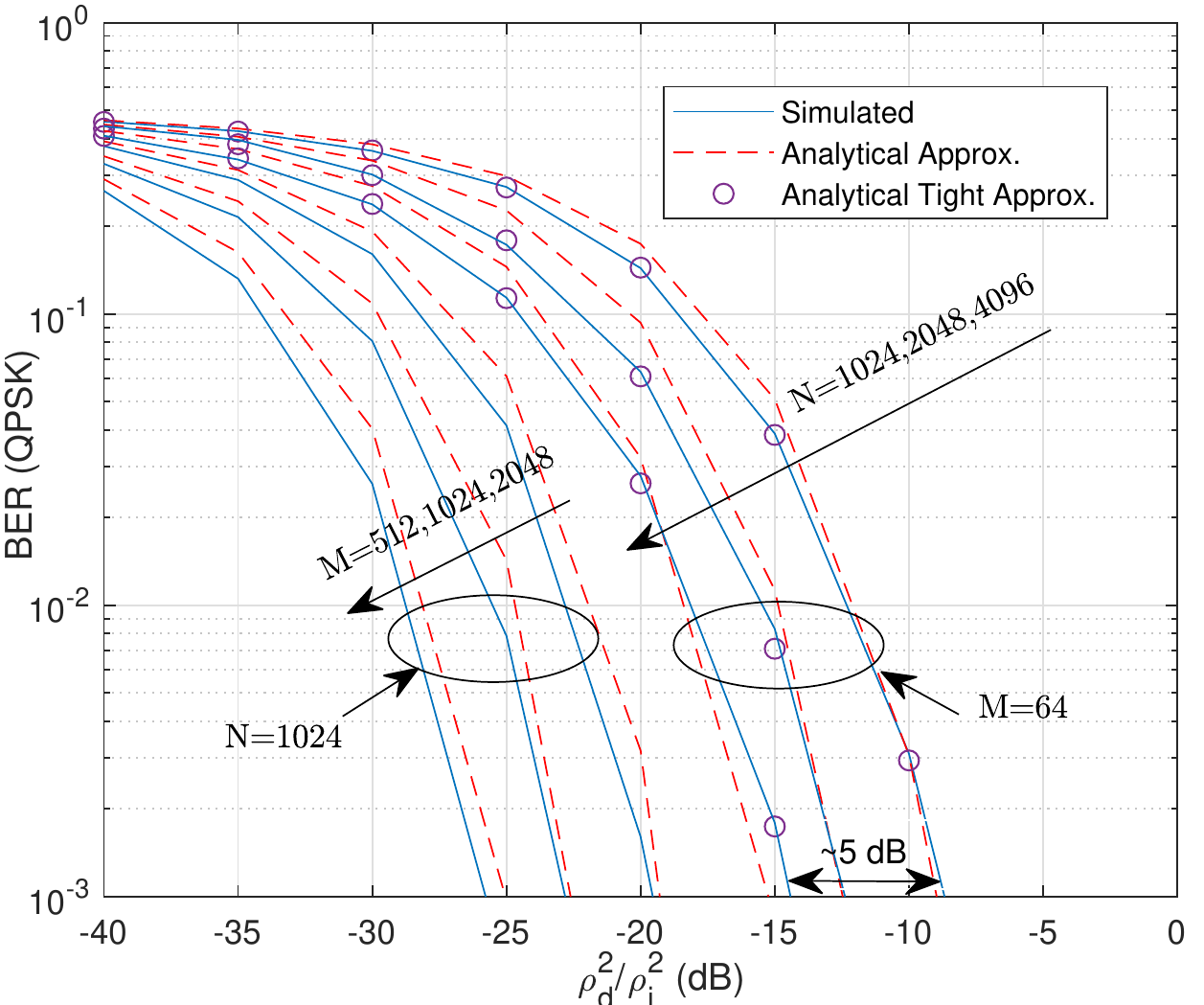}
\caption{BER vs. SIR ($\rho_d^2/\rho_i^2$).}
\label{fig:BER_SIR_var_beta}
\end{subfigure}%
\begin{subfigure}[t]{0.5\textwidth}
\centering
\includegraphics[width=0.98\textwidth]{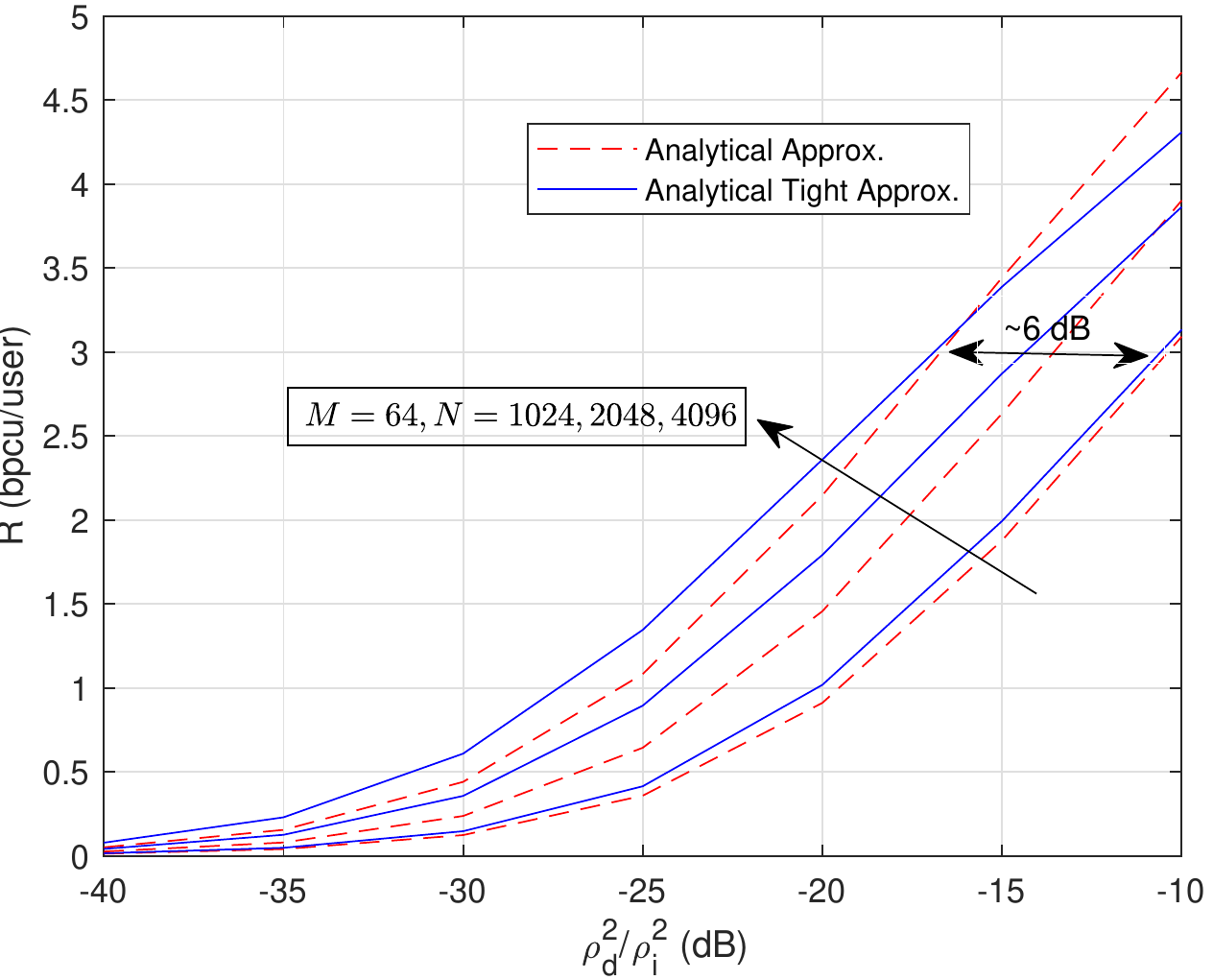}
\caption{$R$ vs. SIR ($\rho_d^2/\rho_i^2$).}
\label{fig:erg_ach_rate}
\end{subfigure}
\caption{BER vs. SIR in (a) and $R$ vs. SIR in (b), $M=64$, $U=I=4$, 1-bit ADC, perfect CSI.}
\end{figure*}

For the simulations, the number of receive antennas is $M=64$, while $U=I=4$. Some other parameters are $N=1024$, $L=10$ and $\mathcal{K}_D=\{N-150, N-149, \ldots, -1, 1, 2, \ldots$ $, 150\}$ ($|\mathcal{K}_D|=300$), while $\mathcal{K}_I=\{250, 251, \ldots, 549\}$ ($|\mathcal{K}_I|=300$), which makes the oversampling rates $\beta_D=\beta_I\approx3.41$. The data symbols are \color{black} quadrature phase-shift keying \color{black}(QPSK) modulated. Subcarrier spacing is 15 KHz as in LTE, with a transmission bandwidth of $|\mathcal{K}_D|/(NT_s)=4.5$ MHz for the desired channel, which does not change with the sampling rate. Furthermore, the noise variance parameter $N_o$ is normalized such that $\rho_d^2/N_o=4$ dB to take $\rho_d^2$ as unity. The type of the multi-bit quantizers is uniform midrise, for which $P_c=1\%$. The power delay profile is taken as uniform, that is, $p[l]=1/L$ for $0\leq\ell<L$. In the plots, the analytical curves for which the SINDR calculation is made based on Proposition~\ref{prop:tight_SINDR} are referred to as ``Analytical Tight Approx.", indicated with dashed lines. Moreover, the curves for which the SINDR is calculated based on Proposition~\ref{prop:approx_SINDR} are named ``Analytical Approx.", indicated with circles in Fig.~\ref{fig:BER_SIR_var_beta} and with solid lines in Fig.~\ref{fig:erg_ach_rate}. As the first case, we change the block length $N$, when all other parameters are fixed (except $L$ which should be directly proportional to $N$) for the perfect CSI condition. The BER vs signal-to-interference ratio (SIR or $\rho_d^2/\rho_i^2$) curves are presented in Fig.~\ref{fig:BER_SIR_var_beta}. \color{black} Note that a single $\rho_d^2/\rho_i^2$ for each data point does not imply that the average received power for every band (desired or interference band) or subcarrier/user is the same for a given channel realization due to (\ref{eqn:freq_dom_eq}).\color{black}
In Fig.~\ref{fig:BER_SIR_var_beta}, the simulated values are indicated with solid lines. As can be noted in the three curves grouped as $M=64$ curves on the right hand side of Fig.~\ref{fig:BER_SIR_var_beta}, the analytical calculations based on Proposition \ref{prop:tight_SINDR} and (\ref{eqn:BER_exp}) are in good agreement with the simulated values. Moreover, the approximate analytical curves based on Proposition \ref{prop:approx_SINDR} generally follow the simulated curves. In addition, we see that increasing the oversampling rate (equivalently increasing the block length while the number of occupied subcarriers is fixed) and the number of antennas $M$ are useful to combat ACI. We can observe up to 5 dB SIR gain by increasing the oversampling rate from $\beta_d\approx\beta_i=3.41 \ (N=1024, L=10)$ to $\beta_d=\beta_i\approx 13.65 \ (N=4096, L=40)$ when the SIR levels to achieve a target BER of $10^{-3}$ is considered. Significant SIR gains is also observed in Fig.~\ref{fig:BER_SIR_var_beta} when $M$ is increased as expected from the analysis.

For the same simulation setting, the ergodic capacity in terms of bits per channel use (bpcu) per user calculated using (\ref{eqn:erg_ach_rate}) are plotted in Fig.~\ref{fig:erg_ach_rate}. An SIR gain more than 5 dB is observed with increasing oversampling rate when the SIR levels to achieve an ergodic capacity of 3 bpcu per user are compared. Moreover, it should be noted that the approximate analytical curve based on Proposition~\ref{prop:approx_SINDR} is close to the tight approximation curve in Proposition~\ref{prop:tight_SINDR} for $N=1024$, a relatively low oversampling rate case, verifying that the approximation in Proposition~\ref{prop:approx_SINDR} is accurate for low oversampling rates and when $L$ is large\footnote{In fact, $R'$ should be compared to a slightly modified version of $R$ in (\ref{eqn:erg_ach_rate}), for which the outer expectation is taken inside the logarithm, but the values obtained for this version are very similar to the values obtained for $R$ in (\ref{eqn:erg_ach_rate}), thus not shown in Fig.~\ref{fig:erg_ach_rate} for the simplicity of the plot.}. \color{black} Furthermore, it can be noticed that the approximate curve based on Proposition 4 yields higher BER for low SIR values or lower BER for high SIR values. The reason for this is as follows. For low SIR values, it can be shown that each element of $\mathbf{C_q}[m]$ will be the samples of an aliased sinc pulse ($m$ being the sample index) centered around $m=0$, all samples being real valued as $\mathcal{K}_D$ is symmetrical around the zeroth subcarrier. Since the values of the tails of the sinc pulse is much lower than that of its main lobe, it is reasonable to assume that $\mathbf{C_q}[m]\approx0$ when $|m|>N/|\mathcal{K}_D|\triangleq W$, as $2N/|\mathcal{K}_D|$ is the null-to-null bandwidth of the sinc pulse). Therefore, $\mathbf{\Gamma}[m]\approx0$ for $|m|>W$. As $\mathbf{C_q}[m]=\mathbf{C_q}[-m]$, $\mathbf{\Gamma}[m]=\mathbf{\Gamma}[-m]$, and $|\mathcal{K}_D|\gg4$,  $\mathrm{DFT}_{N,k}\left\lbrace\mathbf{\Gamma}[m]\right\rbrace\approx\sum_{m=-W}^{W}\mathbf{\Gamma}[m]e^{-j2\pi mk/N}=\sum_{m=-W}^{W}\mathbf{\Gamma}[m]cos(2\pi mk/N)>\mathbf{\Gamma}[0]$. Since the approximation assumes that $\mathbf{\Gamma}[m]=0$ for $m\neq0$ and $\sum_{m=-W}^{W}\mathbf{\Gamma}[m]>\mathbf{\Gamma}[0]$, the quantization noise covariance matrix calculated using Proposition~\ref{prop:error_cov_freq} under this assumption has lower values compared to its exact version, resulting in a higher SINDR calculation than the exact values. For the low SIR case, it can be shown that each element of $\mathbf{C}_{\mathbf{q}[k]}$ will mostly be concentrated inside the interfering band (for $k\in\mathcal{K}_I$) and the quantization noise in the desired band is due to the tails of $\mathrm{sinc^2}
$ pulses, making the variance of the quantization noise in the desired band limited less than the assumed quantization distortion value. The poor performance for the low SIR case is mostly due to the magnitude distortion caused by the matrix $\mathbf{A}$ in (\ref{eqn:no_time_index_quant_sig}).\color{black} 

\begin{figure*}
\centering
\begin{subfigure}[t]{0.5\textwidth}
\centering
\includegraphics[width=0.93\textwidth]{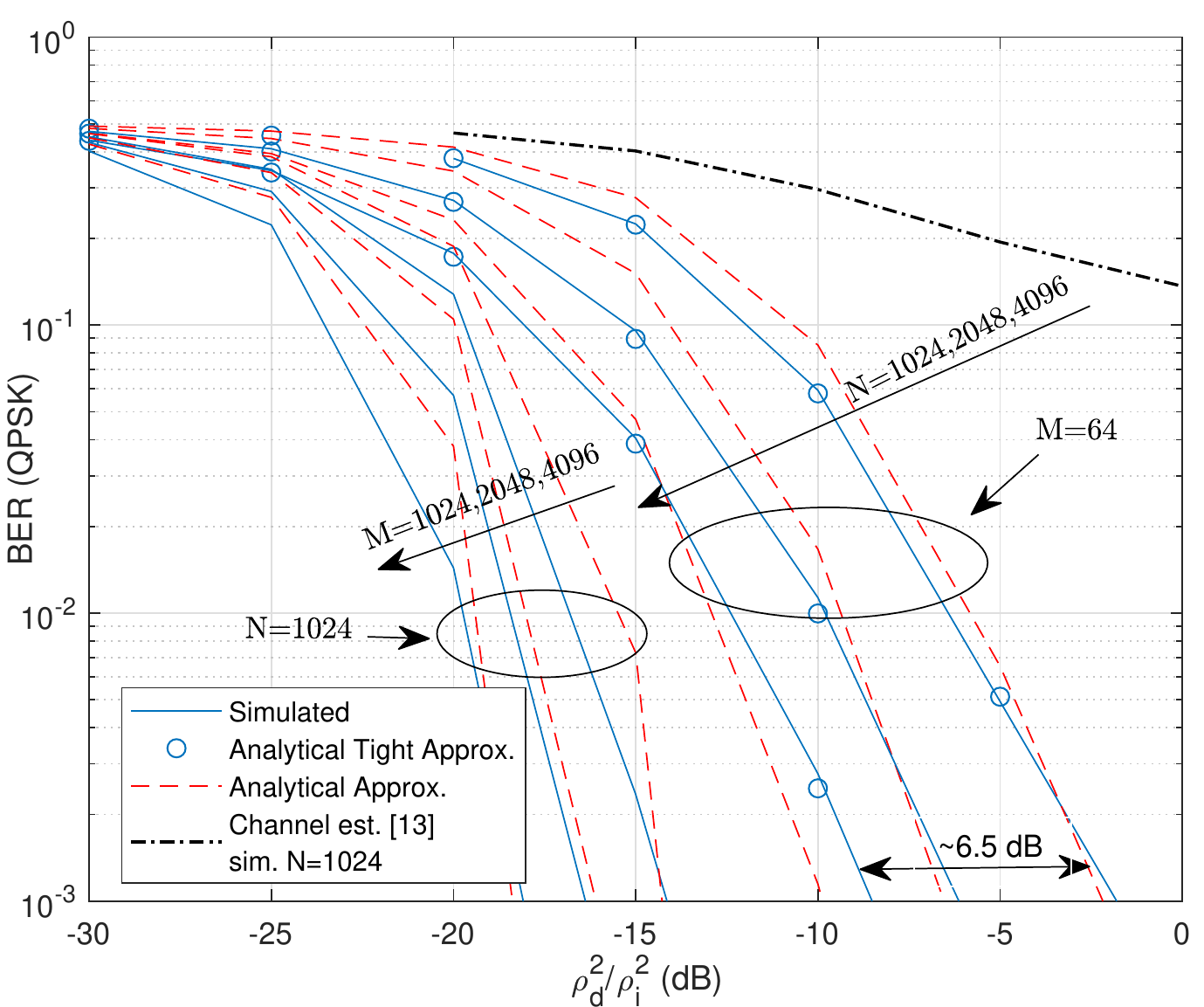}
\caption{BER (QPSK) vs. SIR ($\rho_d^2/\rho_i^2$), one-bit ADC.}
\label{fig:BER_SIR_var_beta_imperf_CSI}
\end{subfigure}%
\begin{subfigure}[t]{0.5\textwidth}
\centering
\includegraphics[width=0.96\textwidth]{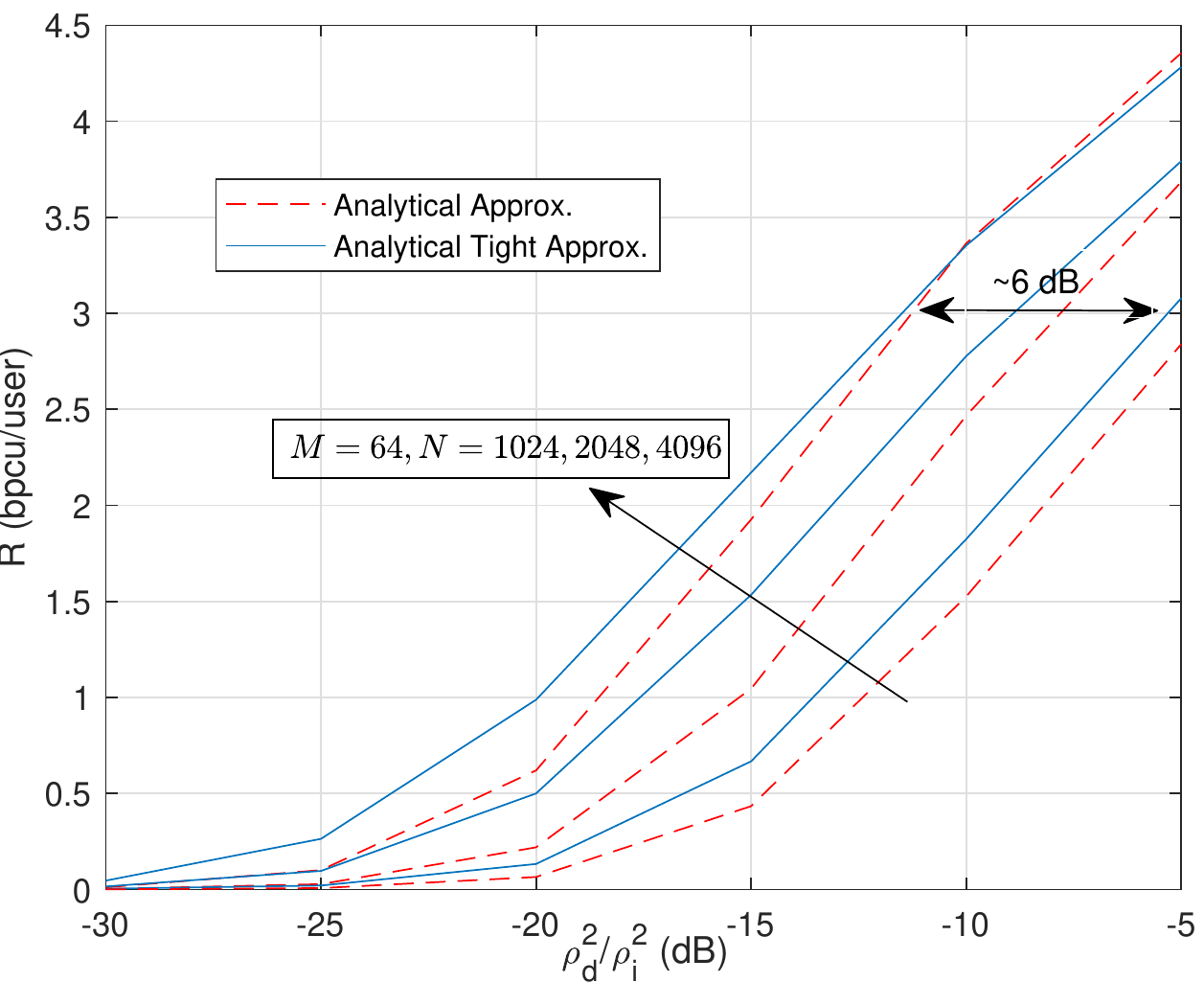}
\caption{$R$ vs. SIR ($\rho_d^2/\rho_i^2$), one-bit ADC.}
\label{fig:erg_ach_rate_impf_CSI}
\end{subfigure}
%\caption{BER vs. SIR (a) and $R$ vs. SIR (b) for $M=64$, $U=I=4$ under imperfect CSI.}
\newline
\begin{subfigure}[t]{0.5\textwidth}
\centering
\includegraphics[width=0.96\textwidth]{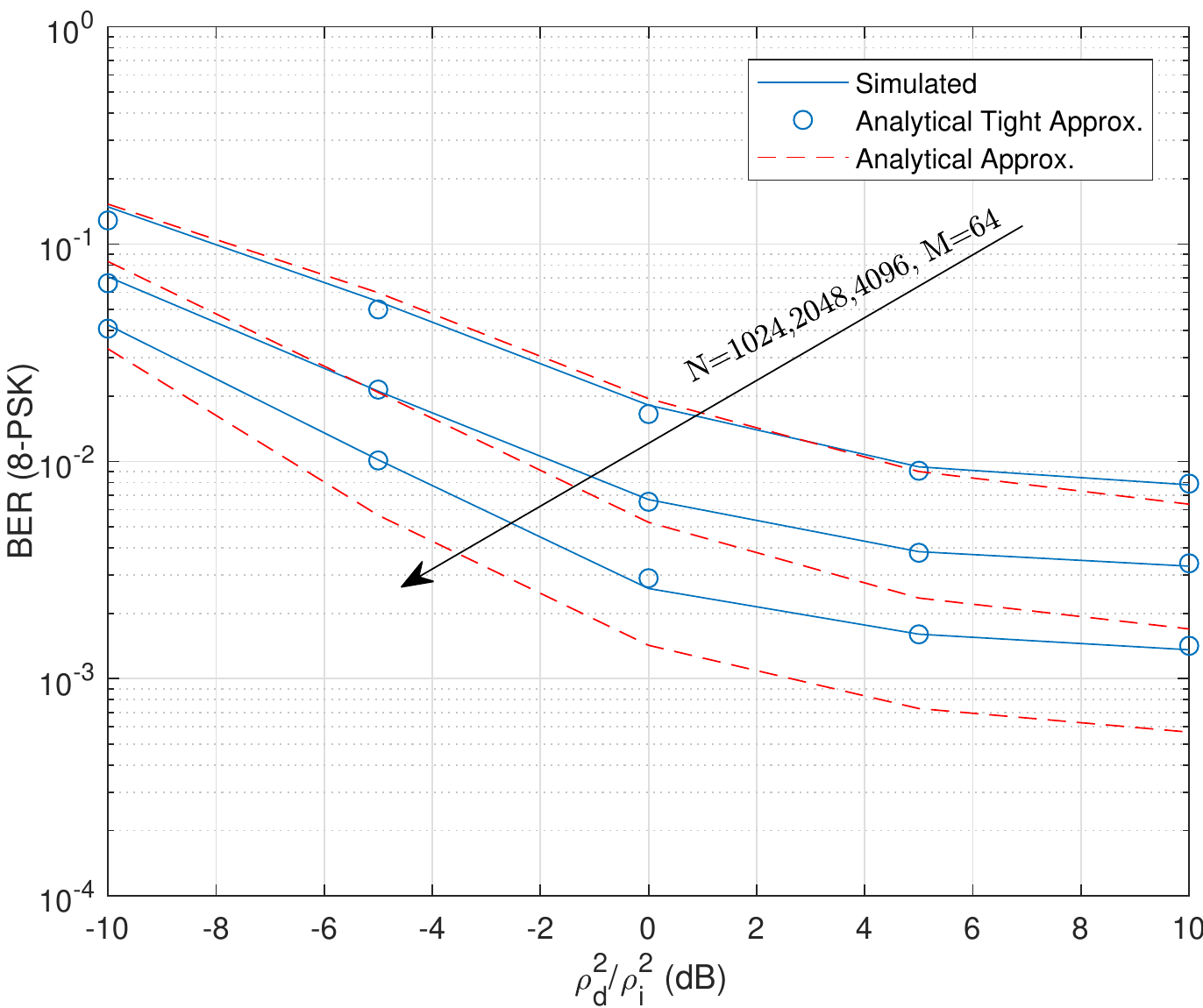}
\caption{BER (8-PSK) vs. SIR ($\rho_d^2/\rho_i^2$), one-bit ADC.}
\label{fig:BER_SIR_var_beta_8PSK}
\end{subfigure}%
\begin{subfigure}[t]{0.5\textwidth}
\centering
\includegraphics[width=0.96\textwidth]{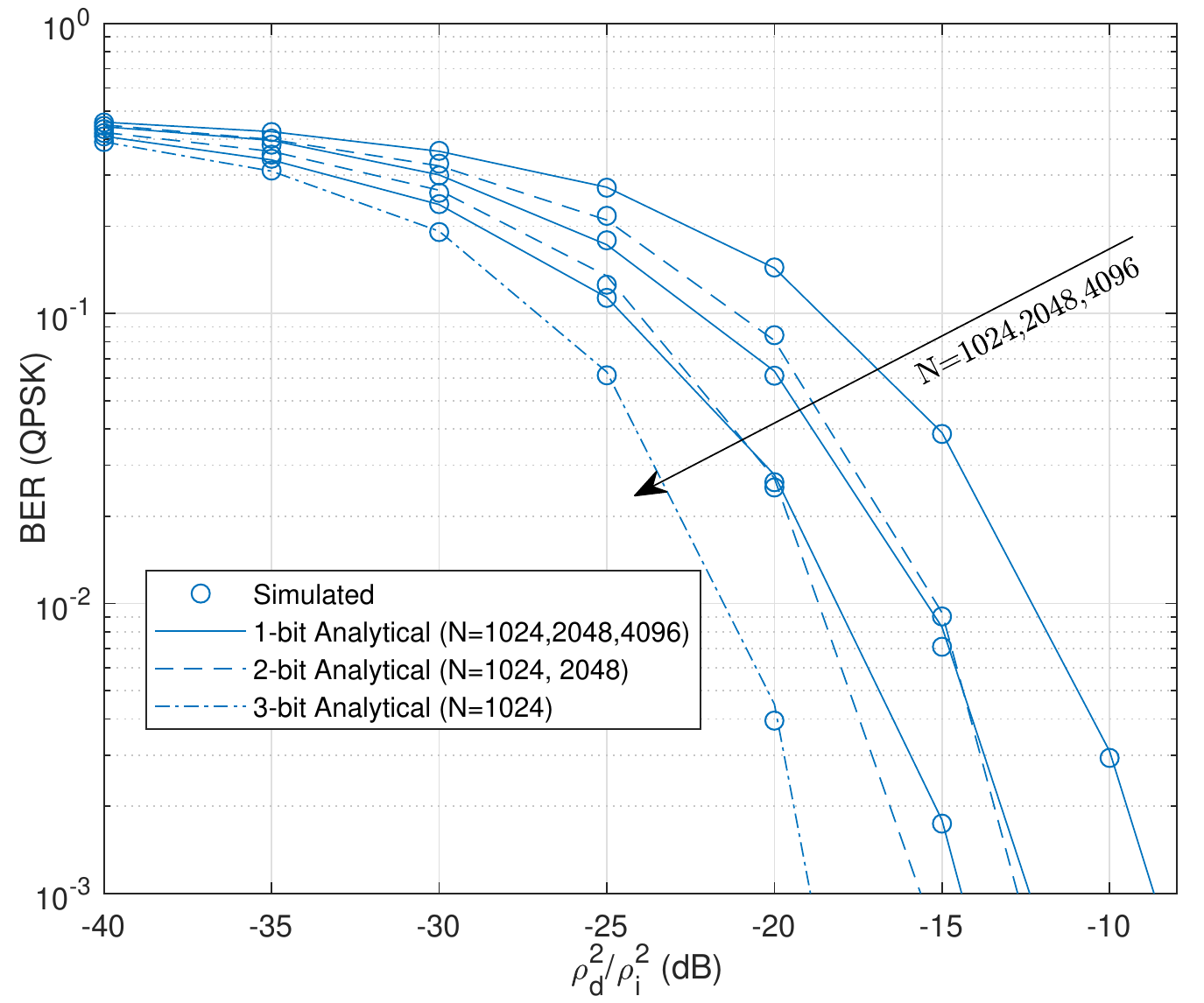}
\caption{BER (QPSK) vs. SIR ($\rho_d^2/\rho_i^2$), multi-bit ADC.}
\label{fig:multi_bit_err_rate_plot}
\end{subfigure}
\caption{Performance plots for imperfect CSI, 1-bit ADC in (a),(b),(c) and multi-bit ADC in (d).}
\end{figure*}

Simulations for the imperfect CSI case are also carried out. The pilot sequence length is taken to be $N$ whereas the set of subcarriers for pilot signals is selected as $\mathcal{K}_P=\mathcal{K}_D\cup\{N-152,N-151,151,152\}$. The noise spectral density $N_o$ in the channel estimation phase is also normalized such that $\rho_p^2/No=4$~dB to take $\rho_p=1$. The BER vs SIR plots are presented in Fig.~\ref{fig:BER_SIR_var_beta_imperf_CSI}.

%\begin{figure}
%\centering
%\includegraphics[width=0.65\columnwidth]{M_64_N__imperf_CSI_all_for_paper_v2.pdf}
%\caption{BER vs. SIR ($\rho_d^2/\rho_i^2$) for $M=64$, $U=I=4$, imperfect CSI case.}
%\label{fig:BER_SIR_var_beta_imperf_CSI}
%\end{figure}
As can be noted in Fig.~\ref{fig:BER_SIR_var_beta_imperf_CSI}, the analytical BER curves obtained based on the SINDR calculation in Proposition~\ref{prop:tight_SINDR} are very close to the simulated results. Moreover, while the approximate analytical curve is not as close to the simulated values as the "Analytical Tight Approx." curves, it can follow the error rate curves in general. Moreover, it can also be deduced from Fig.~\ref{fig:BER_SIR_var_beta_imperf_CSI} that the 5 dB SIR gain achieved with oversampling for the perfect CSI case can also be attained under imperfect channel knowledge when the error rates to achieve a BER value of $10^{-3}$ is considered. In fact, it is even more than 5 dB (about 6.5 dB) for imperfect CSI. This is because oversampling also enhances channel estimation quality which in turn results in a better error rate performance even further for the one-bit quantized system with imperfect CSI. Comparing the perfect and imperfect CSI BER curves in Fig.~\ref{fig:BER_SIR_var_beta} and Fig.~\ref{fig:BER_SIR_var_beta_imperf_CSI}, we observe about 6.5 dB SIR loss due to channel estimation error. This is not an unexpected value, as it is known that the normalized channel estimation \color{black} mean squared error \color{black}converges to -4.4 dB for infinite training power with one-bit ADCs \cite{Li_perf_analysis}, resulting in a similar SIR loss according to (\ref{eqn:approx_SINDR_expression}). The remaining 2.1 dB loss is due to the finite training power, which is close to the SNR loss of about 2 dB for a MIMO setting with infinite ADC resolution \cite{imperf_csi_SNR_loss_mimo}. \color{black} Moreover, to compare the proposed channel estimation algorithm with an existing method of comparable complexity in \cite{Mollen_wideband}, which neither considers the effect of ACI nor employs oversampling for channel estimation, we made simulations to obtain the BER performance of our system with $N=1024$ which uses the channel estimates obtained with the channel estimation method in \cite{Mollen_wideband}. The corresponding BER curve is labeled as ``Channel Est. [13] Sim. N=1024" in Fig.~\ref{fig:BER_SIR_var_beta_imperf_CSI}. As can be noted, a significant performance loss is observed if the channel estimation method in \cite{Mollen_wideband} is used instead of our method.\color{black}

In addition to the BER curves, the ergodic capacity curves for imperfect CSI are also presented in Fig.~\ref{fig:erg_ach_rate_impf_CSI}.
%\begin{figure}
%\centering
%\includegraphics[width=0.65\columnwidth]{N_all_ach_rate_imperf_CSI_analyt_v2.pdf}
%\caption{$R$ vs. SIR ($\rho_d^2/\rho_i^2$) for $M=64$, $U=I=4$ under imperfect CSI.}
%\label{fig:erg_ach_rate_impf_CSI}
%\end{figure}
As can be noted in Fig.~\ref{fig:erg_ach_rate_impf_CSI}, more than 5 dB SIR advantage can be obtained with temporal oversampling when the SIR levels to achieve a target ergodic rate per user value of 3 bpcu/user are compared. Moreover, the ``analytical approximate" curve can generally follow the ``tight approximation" curve. The reason for the approximate ergodic capacity curve is not as close to the tight approximation curve for $N=1024$ as in the perfect CSI case is due to the additional approximation error due to the assumptions involved in the proposed channel estimation scheme, distorting the orthogonality of the channel estimates and the estimation errors.

\color{black}Regarding the performance with higher-order modulations, BER vs. SIR plots for 8-PSK and imperfect CSI are presented in Fig.~\ref{fig:BER_SIR_var_beta_8PSK}. As can be expected, worse BER performance is observed compared to QPSK. An error floor is observed since the quantization and thermal noise exist even if the interference power is zero (infinite SIR). Moreover, a significant BER performance advantage is obtained with oversampling. The tight approximation based on Proposition \ref{prop:tight_SINDR} closely approximates the simulated values while the approximate curves based on Proposition 4 provides accurate values for low oversampling rates as expected from the discussion in Appendix \ref{apdx:prop4_proof}. \color{black}

The error rate curves are also plotted for multi-bit quantizers (up to 3-bits) in Fig.~\ref{fig:multi_bit_err_rate_plot}.
%\begin{figure}
%\centering
%\begin{subfigure}[t]{0.5\textwidth}
%\includegraphics[width=0.95\textwidth]{M_64_N__imperf_CSI_all_for_paper_8PSK.pdf}
%\caption{BER (8-PSK) vs. SIR ($\rho_d^2/\rho_i^2$).}
%\label{fig:BER_SIR_var_beta_8PSK}
%\end{subfigure}%
%\begin{subfigure}[t]{0.5\textwidth}
%\includegraphics[width=0.91\textwidth]{M_64_N_all_for_multi_bit_new_v2.pdf}
%\caption{BER (QPSK) vs. SIR ($\rho_d^2/\rho_i^2$), multi-bit quantizer.}
%\label{fig:multi_bit_err_rate_plot}
%\end{subfigure}
%\caption{BER vs. SIR for 8-PSK, one-bit quantizer, imperfect CSI (a), QPSK, multi-bit quantizer, perfect CSI for $M=64$, $U=I=4$ (b).}
%\end{figure}
%
%
%%\begin{figure}
%%\centering
%%\includegraphics[width=0.55\columnwidth]{M_64_N_all_for_multi_bit_new_v2.pdf}
%%\caption{BER vs. SIR ($\rho_d^2/\rho_i^2$) for $M=64$, $U=I=4$, multi-bit quantizer.}
%%\label{fig:multi_bit_err_rate_plot}
%%\end{figure}
As can be noted in Fig.~\ref{fig:multi_bit_err_rate_plot}, the simulated values are very close to the analytical BER curves which are based on the SINDR calculation in Proposition~\ref{prop:tight_SINDR}. In Fig.~\ref{fig:multi_bit_err_rate_plot}, the oversampling rate increases from $1024$ to $4096$ ($L$ from $10$ to $40$) towards left for all quantization resolutions (1 bit to 3 bits). However, it should be noted that for 2-bit quantizer, there are two cases, either $N=1024$ and $N=2048$, while there is only the BER curve for $N=1024$ for 3-bit quantizer. \color{black} As can also be noted in Fig.~\ref{fig:multi_bit_err_rate_plot}, the simulated values are in perfect agreement with the analytical values, even for the 2-bit quantizer, thus, it can be stated that the assumption of uncorrelated quantization noise in time is accurate unless the ADC resolution is as low as one bit (the quantization noise correlation in time is taken into account for the one-bit ADC case).\color{black}

In order to make fair comparisons between the cases presented in Fig.~\ref{fig:multi_bit_err_rate_plot}, we will try to equate the power consumptions of the various quantization resolution and oversampling rate cases. It is assumed that the power consumption of an ADC is proportional to $2^q$, that is, the power consumption is doubled for single bit addition. This assumption is verified to be accurate in various studies \cite{murmann_extra_dB,
ADCsurveyandAnalysis}. It is also assumed that ADC power consumption grows linearly with oversampling rate as in \cite{ADCsurveyandAnalysis}.

Equating the power consumptions, the first cases to be compared are 1-bit ADC with $N=2048$ and 2-bit ADC with $N=1024$. As can be noted in Fig.~\ref{fig:multi_bit_err_rate_plot}, ACI supression in the 1-bit ADC with $N=2048$ case is better compared to the 2-bit ADC with $N=1024$ case for BER values higher than $10^{-2}$, while the 2-bit ADC with $N=1024$ case is slightly better for BER values lower than $10^{-3}$. Therefore, it can be stated that 1-bit ADC with $N=2048$ is preferable over a 2-bit ADC with $N=1024$, as it provides better BER values and the complexity of a 1-bit ADC is significantly lower (there is no need for an AGC unit in a 1-bit ADC, die area is doubled with every bit increase for flash ADCs; and component matching requirements are also doubled with every bit increase for flash, sucessive approximation (SAR) or pipelined ADCs \cite{sumathi2007labview}). Moreover, the time it takes to complete a conversion (conversion time) is also doubled with every bit of increase for integrating ADCs, while the conversion time scales linearly with number of bits for SAR or pipelined converters \cite{sumathi2007labview}. Therefore, by oversampling with a 1-bit ADC, while we achieve better error rate performance than 2-bit ADCs, total power consumptions equated, we also have advantages regarding ADC complexity and conversion time.

We can also compare other two cases, one is the performance of 1-bit ADC with $N=4096$ and the other is 2-bit ADC with $N=2048$, as their power consumptions are equal. We see from Fig.~\ref{fig:multi_bit_err_rate_plot} that their performances are nearly equal for BER values lower than $10^{-2}$, while the 2-bit ADC with $N=2048$ case has about 1.5 dB SIR advantage for the BER value of $10^{-3}$. The design engineer should be considering whether it is worth to have $1.5$ dB SIR advantage to use 2-bit ADCs, which have the aforementioned disadvantages regarding implementation complexity and conversion time compared to the 1-bit ADCs.

The remaining performance comparison is between 3-bit ADCs with $N=1024$ and 2-bit ADCs with $N=2048$, as their power consumptions are equal. We can see that we have about 4 dB SIR advantage with 3-bit ADC. However, again, it should also be considered that such an SIR gain does not come for free, as 2-bit ADCs are much more advantageous compared to 3-bit ADCs in terms of die area, component matching circuitry and conversion time, thus 2-bit ADCs with oversampling can be a choice compared to 3-bit ADCs despite the SIR disadvantage.

\section{Conclusions}
In this paper, we have presented a performance analysis for an uplink massive MIMO-OFDM system with low-resolution oversampling ADCs under frequency selective fading in an interfering adjacent channel interference scenario for perfect or imperfect receiver CSI. The analysis arrived at two important expressions, one of which gives very accurate results but limited insights, the other giving noticeable approximation errors but much clearer insights into the dependence of system performance on the system parameters. We have shown both with analysis and simulations that adjacent band interference can be suppressed by increasing the number of antennas or the oversampling rate. Moreover, we discussed whether to use lower-resolution ADCs with higher oversampling rates or higher-resolution ADCs with lower oversampling rates comparing their error rate performances while their power consumptions are equated.

%Therefore, an alternative method of lower complexity to compute $\mathbf{A}$ will be presented. Let $\mathbf{C}_{\text{\ul{$\mathbf{r}$}}}[n]$, $n=1,\cdots,N$ be a diagonal matrix whose diagonal elements are equal to a set of diagonal elements of matrix $\mathbf{C}_{\text{\ul{$\mathbf{r}$}}}$, which are $\left[(\mathbf{C}_{\text{\ul{$\mathbf{r}$}}})_{n,n} \ (\mathbf{C}_{\text{\ul{$\mathbf{r}$}}})_{n+1,n+1} \cdots \ (\mathbf{C}_{\text{\ul{$\mathbf{r}$}}})_{n+M-1,n+M-1} \right]$. Since it can be derived using (\ref{eqn:rec_sig_discr_time_mtx_vec}) that $\mathbf{r}[n]$ a stationary process, that is, $\mathbb{E}\left[\mathbf{r}[n]\mathbf{r}[n-m]^H\right]$ does not depend on $n$, it can be shown that $\mathbf{C}_{\text{\ul{$\mathbf{r}$}}}[n]=\mathbf{C}_{\text{\ul{$\mathbf{r}$}}}[n']$ for $n\neq n'$.
%This implies that $\mathbf{A}[n]=\mathbf{A}[n']$ for $n\neq n'$, where $\mathbf{A}[n]=\left[(\mathbf{A})_{n,n} (\mathbf{A})_{n+1,n+1} \cdots \ (\mathbf{A})_{n+M-1,n+M-1} \right]$. With these definitions along with (\ref{eqn:a_mtx_exp}), it can be written that
%\begin{equation}
%\label{eqn:mtx_a_sub}
%\mathbf{A}[n]=\sqrt{\dfrac{4}{\pi}}\mathrm{diag}(\mathbf{C}_{\text{\ul{$\mathbf{r}$}}}[n])^{-0.5}
%\end{equation}
%By definition $\mathbf{C}_{\text{\ul{$\mathbf{r}$}}}[n]=$

\appendices
\section{Proof of Proposition 1}
\label{apdx:prop1_proof}
Direct computation yields
\begin{align}
    \mathbf{C}_{\mathbf{r}}[m]=\mathbb{E}\left[\mathbf{r}[n]\mathbf{r}[n-m]^H\right]&=\mathbb{E}\left[\dfrac{1}{N^2}\sum_{k=0}^{N-1}\sum_{k'=0}^{N-1}\tilde{\mathbf{r}}[k]\tilde{\mathbf{r}}[k']^He^{j2\pi nk/N}e^{-j2\pi (n-m)k'/N}\right] \label{eqn:prop_prf_l1}\\
    &=\dfrac{1}{N^2}\sum_{k=0}^{N-1}\mathbb{E}\left[\tilde{\mathbf{r}}[k]\tilde{\mathbf{r}}[k]^H\right]e^{j2\pi nk/N}e^{-j2\pi (n-m)k/N}
    \label{eqn:prop_prf_l3}\\
    &=\dfrac{1}{N^2}\sum_{k=0}^{N-1}\mathbf{C}_{\mathbf{\tilde{r}}[k]}e^{j2\pi mk/N}\label{eqn:prop_prf_l4}.
\end{align}
Here (\ref{eqn:prop_prf_l1}) holds by definition, (\ref{eqn:prop_prf_l3}) is because $\mathbb{E}\left[\tilde{\mathbf{r}}[k]\tilde{\mathbf{r}}[k']^H\right]=\mathbf{0}$ for $k\neq k'$ as the data symbols of the users are independent and (\ref{eqn:prop_prf_l4}) also holds by definition.\hfill \qedsymbol
\section{Proof of Proposition 2}
\label{apdx:prop2_proof}
$\mathbf{C}_{\mathbf{q}[k]}$ \text{can be found from} $\mathbf{C_q}[m]$ as
\begin{align}
    \mathbf{C}_{\mathbf{q}[k]}=\mathbb{E}\left[\mathbf{q}[k]\mathbf{q}[k]^H\right]&=\sum_{m=0}^{N-1}\sum_{m'=0}^{N-1}\mathbb{E}\left[\tilde{\mathbf{q}}[m]\tilde{\mathbf{q}}[m']^H\right]e^{-j2\pi mk/N}e^{j2\pi m'k/N}\label{eqn:prop2_prf_l2}\\
    &=\sum_{m=0}^{N-1}\sum_{m'=0}^{N-1}\mathbf{C}_{\mathbf{q}}[m-m']e^{-j2\pi (m-m')k/N}\label{eqn:prop2_prf_l3}\\
    &=\sum_{\ell=-(N-1)}^{N-1}(N-|\ell|)
    \mathbf{C}_{\mathbf{q}}[\ell]
    e^{-j2\pi \ell k/N}\label{eqn:prop2_prf_l4}\\
    &=\left[\sum_{\ell=0}^{N-1}(N-\ell)
    \mathbf{C}_{\mathbf{q}}[\ell]
    e^{-j2\pi \ell k/N}+\sum_{\ell=0}^{N-1}(N-\ell)\mathbf{C}_{\mathbf{q}}[\ell]^{*}
    e^{j2\pi \ell k/N}\right]\label{eqn:prop2_prf_l5}\\
    &=\mathrm{DFT}_{N,k}\left\lbrace\mathbf{\Gamma}[\ell]\right\rbrace+\mathrm{DFT}_{N,k}\left\lbrace\mathbf{\Gamma}[\ell]\right\rbrace^*-N\mathbf{C_{\tilde{q}}}[0].\label{eqn:prop2_prf_l6}
\end{align}

Here (\ref{eqn:prop2_prf_l2})-(\ref{eqn:prop2_prf_l3}) hold by definition and due to stationarity of $\mathbf{q}[m]$. A change of variable ($\ell=m-m'$) is introduced in (\ref{eqn:prop2_prf_l4}). (\ref{eqn:prop2_prf_l5}) holds since $\mathbf{C}_{\mathbf{q}}[\ell]^{*}=\mathbf{C}_{\mathbf{q}}[-\ell]$.
\hfill \qedsymbol
\section{Proof of Proposition 3}
\label{apdx:prop3_proof}
When $\gamma_u[k]$ is as defined in (\ref{eqn:sinr_cond_h_est}), the fact that $R$ in (\ref{eqn:erg_ach_rate}) is a lower bound ergodic capacity follows from \cite[eq. (2.46)]{Marzetta1088544}, since the conditions $\mathbb{E}[w_u[k]|\mathbf{\hat{H}}]=\mathbb{E}[\tilde{s}_u[k]^*w_u[k]|\mathbf{\hat{H}}]=\mathbb{E}[\hat{g}_u[k]'\tilde{s}_u[k]^*w_u[k]|\mathbf{\hat{H}}]=0$, where $w_u[k]\triangleq i_u[k]+n_u[k]+q_u[k]$, hold for LMMSE channel estimate $\mathbf{\hat{H}}$. $\mathbb{E}[g_u[k]'|\mathbf{\hat{H}}]$ in the numerator of $\gamma_u[k]$ expression can be found as follows:
%Conditioned on $\mathbf{\hat{H}}$, $\mathbf{\hat{B}}[k]\triangleq\left(\mathbf{{\hat{H}}}[k]^H\mathbf{{\hat{H}}}[k]\right)^{-1}\mathbf{{\hat{H}}}[k]^H$ will be deterministic. This implies that $\mathbf{\hat{b}}_u[k]$ can be considered to be deterministic as well as $\hat{\mathbf{h}}_u[k]$. Moreover, the conditional mean of $n_u[k]$ and $q_u[k]$ are also zero as thermal noise $\mathbf{w}[k]$ and the quantization noise $\mathbf{q}[n]$ are zero-mean. The conditional mean of $i_u[k]$ is also zero since $s_z[k]$, $\forall z\neq u$ are zero mean and LMMSE estimation error is unbiased.
\begin{align}
\label{eqn:gain_mean}
\mathbb{E}\left[g_u[k]'|\mathbf{\hat{H}}\right]&=\mathbb{E}\left[\rho_d\sqrt{N}\mathbf{\hat{b}}_u[k]^H\mathbf{A}{\mathbf{\tilde{h}}_u}[k]||\mathbf{\hat{H}}\right]\\
&=\mathbb{E}\left[\rho_d\sqrt{N}\mathbf{\hat{b}}_u[k]^H\mathbf{A}{\mathbf{\hat{h}}_u}[k]||\mathbf{\hat{H}}\right]+\mathbb{E}\left[\rho_d\sqrt{N}\mathbf{\hat{b}}_u[k]^H\mathbf{A}{\mathbf{{\tilde{e}}}_u}[k]||\mathbf{\hat{H}}\right]\\
&=\rho_d\sqrt{N}\mathbf{\hat{b}}_u[k]^H\mathbf{A}{\mathbf{\hat{h}}_u}[k],
\end{align}
where the last step is owing to the fact that LMMSE channel estimates are unbiased thus the estimation errors are zero-mean.
%\begin{align}
%\mathbb{E}[i_u[k]|&\mathbf{\hat{H}},\tilde{s}_u[k]]=\mathbf{\hat{b}}_u[k]^H\mathbf{A}{\mathbf{\tilde{e}}_u}[k]\tilde{s}_u[k].
%\label{eqn:unbiased_est}
%\end{align}
$\text{Var}\left[\hat{g}_u[k]'|\mathbf{\hat{H}}\right]$ can also be found as follows:
\begin{align}
\text{Var}\left[\hat{g}_u[k]'|\mathbf{\hat{H}}\right]&=\text{Var}\left[\rho_d\sqrt{N}\mathbf{\hat{b}}_u[k]^H\mathbf{A}{\mathbf{\hat{h}}_u}[k]+\rho_d\sqrt{N}\mathbf{\hat{b}}_u[k]^H\mathbf{A}{\mathbf{\tilde{e}}_u}[k]|\mathbf{\hat{H}}\right]\\
&=\text{Var}\left[\rho_d\sqrt{N}\mathbf{\hat{b}}_u[k]^H\mathbf{A}{\mathbf{\tilde{e}}_u}[k]|\mathbf{\hat{H}}\right]\\
&=\rho_d^2N{\mathbf{\hat{b}}_u}[k]^H\mathbf{A}\mathbb{E}\left[\mathbf{\tilde{e}}_u[k]\mathbf{\tilde{e}}_u[k]^H\right]\mathbf{A}^H\mathbf{\hat{b}}_u[k]=\rho_d^2N\sigma_e^2||\mathbf{\hat{b}}_u[k]^H\mathbf{A}||^2\label{eqn:uncor_chan_err_wh_chan_est},
\end{align}
where (\ref{eqn:uncor_chan_err_wh_chan_est}) follows from the fact that the LMMSE channel estimation errors are uncorrelated with the channel estimates, hence with the ZF matrix row vectors.
Moreover,
\begin{align}
\text{Var}[i_u[k]+n_u[k]+q_u[k]|\mathbf{\hat{H}}]=&\text{Var}\left[i_u[k]|\mathbf{\hat{H}}\right]+\text{Var}\left[n_u[k]|\mathbf{\hat{H}}\right]+\text{Var}\left[q_u[k]|\mathbf{\hat{H}}\right].\label{eqn:indep_var}
\end{align}
(\ref{eqn:indep_var}) is due to uncorrelatedness of the thermal and quantization noise with every other term. The uncorrelatedness of the quantization noise with the other terms is due to the Bussgang decomposition; matrix $\mathbf{A}$ is selected to make $\mathbf{r}[n]$ to be uncorrelated with $\mathbf{q}[n]$ in (\ref{eqn:no_time_index_quant_sig}) and this also holds for the frequency domain terms as DFT is a unitary transformation which preserves inner products. Moreover, $\text{Var}\left[i_u[k]|\mathbf{\hat{H}}\right]$ can be derived as follows:
\begin{align}
\text{Var}&\left[i_u[k]|\mathbf{\hat{H}}\right]=\rho_d^2N\text{Var}\left[\sum_{z\neq u, z\in\mathcal{U}_d}\mathbf{{\hat{b}}}_u[k]^H\mathbf{A}{\mathbf{\hat{h}}_z}[k]\tilde{s}_z[k]+\sum_{z\neq u, z\in\mathcal{U}_d}{\mathbf{\hat{b}}_u}[k]^H\mathbf{A}\mathbf{\tilde{e}}_z[k]\tilde{s}_z[k]\biggr\rvert\mathbf{\hat{H}}\right]\label{eqn:var_int_zero_mean}\\
&=\rho_d^2N\left[\sum_{z\neq u, z\in\mathcal{U}_d}|\mathbf{{\hat{b}}}_u[k]^H\mathbf{A}{\mathbf{\hat{h}}_z}[k]|^2+\sum_{z\neq u, z\in\mathcal{U}_d}{\mathbf{\hat{b}}_u}[k]^H\mathbf{A}\mathbb{E}\left[\mathbf{\tilde{e}}_z[k]\mathbf{\tilde{e}}_z[k]^H\right]\mathbf{A}^H\mathbf{\hat{b}}_u[k]\right]\label{eqn:var_int_zero_mean2}\\
&=\rho_d^2N\left[\sum_{z\neq u, z\in\mathcal{U}_d}|\mathbf{{\hat{b}}}_u[k]^H\mathbf{A}{\mathbf{\hat{h}}_z}[k]|^2+\sum_{z\neq u, z\in\mathcal{U}_d}{\mathbf{\hat{b}}_u}[k]^H\mathbf{A}\sigma_e^2\mathbf{A}^H\mathbf{\hat{b}}_u[k]\right]\label{eqn:var_int_zero_mean3}\\
&=\rho_d^2N\left[\sum_{z\neq u, z\in\mathcal{U}_d}|\mathbf{{\hat{b}}}_u[k]^H\mathbf{A}{\mathbf{\hat{h}}_z}[k]|^2+\sigma_e^2(U-1)||{\mathbf{\hat{b}}_u}[k]^H\mathbf{A}||^2\right]\label{eqn:var_int_zero_mean4}
\end{align}
where (\ref{eqn:var_int_zero_mean2}) is follows from the fact that the channel estimation errors for all users are also uncorrelated with each other and with ZF matrix row vectors. Moreover, (\ref{eqn:var_int_zero_mean3}) is due owing to the fact that estimation errors for all channel coefficients are the same due to (\ref{eqn:analyt_est_err}). It is also straightforward to show that $\text{Var}\left[n_u[k]|\mathbf{\hat{H}},\tilde{s}_u[k]\right]=N_u[k]$, $\text{Var}\left[q_u[k]|\mathbf{\hat{H}},\tilde{s}_u[k]\right]=Q_u[k]$. \hfill \qedsymbol

\section{Proof of Proposition 4}
\label{apdx:prop4_proof}

Denote the $m^{th}$ element of $\mathbf{C_r}[0]$ by $\mathbf{C_r}[0]_{m,m}$, which can be calculated as
\begin{align}
\label{eqn:mean_cond_H}
\mathbf{C_r}[0]_{m,m}&=\mathbb{E}\left[\left\lvert\sum_{\ell=0}^{L-1}\sum_{u=1}^{U+I}h_{m,u}[\ell]s_u[n-l]+w_m[n]\right\rvert^2\bigg\rvert\mathbf{\underline{H}}\right]=\sum_{\ell=0}^{L-1}\sum_{u=1}^{U+I}G_u|h_{m,u}[\ell]|^2+N_o,
\end{align}
$G_u=\left(|\mathcal{K}_D|\rho_d^2\right)/N$ if $u\in\mathcal{U}_D$ or $G_u=\left(|\mathcal{K}_I|\rho_i^2\right)/N$ if $u\in\mathcal{U}_I$. Owing to Chebyshev inequality,
\begin{equation}
\label{eqn:chebyshev_ineq}
\mathrm{Pr}\left[\left\lvert\mathbf{C_r}[0]_{m,m}-\mathbf{\mu}\right\rvert\geq{\epsilon}\right]\leq\frac{\mathrm{Var}[\mathbf{C_r}[0]_{m,m}]}{\epsilon^2} \text{ }   \forall m,
\end{equation}
where $\mathbf{\mu}\triangleq \mathbb{E}_{\mathbf{\underline{H}}}\left[\mathbf{C_r}[0]_{m,m}\right]=\left(|\mathcal{K}_D|U\rho_d^2+|\mathcal{K}_I|I\rho_i^2\right)/N+N_o$, $\mathrm{Pr}[\xi]$ denote the probability of an event $\xi$ and $\mathrm{Var}[\mathbf{C_r}[0]_{m,m}]$ is the variance of $\mathbf{C_r}[0]_{m,m}$ with respect to $\mathbf{\underline{H}}$, which is equal to $3G_u^2/L$. Since $\mathrm{Var}[\mathbf{C_r}[0]_{m,m}]$ is decreasing with $L$, for any $\epsilon>0$ we can find $L>0$ such that $\mathrm{Pr}\left[\left\lvert\mathbf{C_r}[0]_{m,m}-\mathbf{\mu}\right\rvert\geq{\epsilon}\right]=0$, thus $\mathbf{C_r}[0]_{m,m}$ converge in probability to $\mathbf{\mu}$. Therefore, the error in the approximation $\mathbf{C_r}[0]_{m,m}\approx\left(|\mathcal{K}_D|U\rho_d^2+|\mathcal{K}_I|I\rho_i^2\right)/N+N_o$ converge to zero in probability as $L$ grows large (goes to infinity). %Therefore, $\mathrm{Var}\{\mathbf{C_r}[0]_{m,m}\}$ is the variance of $\mathbf{C_r}[0]_{m,m}$. Define normalized error ${\epsilon'}\triangleq\epsilon/\mu$ with respect to $\mu$. It can be shown that $\mathrm{Var}\{\mathbf{C_r}[0]_{m,m}\}=3G_c(U+I)/L$ and $\mu=(U+I)G_c+N_o$, thus $\mu^2=(U+I)^2G_c^2+N_o^2+2(U+I)G_c$. Therefore, (\ref{eqn:chebyshev_ineq}) implies that  As $\mathrm{Var}\{\mathbf{C_r}[0]_{m,m}\}$ does not change with $L$ It follows from (\ref{eqn:chebyshev_ineq}) that for every  By selecting $L$ or $U+I$ large enough, we can decrease the normalized error $\epsilon'$ it can be shown that while $\mathrm{Var}\mathbf{C_r}[0]_{m,m}$ increase linearly with $L$ or $U+I$, since ${\epsilon'}$ increase with $L^2$ and $(U+I)^2$, the normalized error goes to zero as $L$ or $(U+I)$ goes large.
This implies that
\begin{align}
\mathbf{A}&=\dfrac{2}{\sqrt{\pi}}\mathrm{diag}\left(\mathbf{C_r}[0]\right)^{-0.5}\rightarrow\dfrac{2}{\sqrt{\pi}}\left(\left(|\mathcal{K}_D|U\rho_d^2+|\mathcal{K}_I|I\rho_i^2\right)/N+N_o\right)^{-0.5}\mathbf{I}=G\mathbf{I},\label{eqn:rec_pow}
\end{align}
as $L$ grows large (convergence is represented by $\rightarrow$ symbol). It can similarly be shown that $\mathbf{C_r}[0]_{m,i}\approx 0$ for $m\neq i$ as $L$ grows large since $\mathbb{E}_{\mathbf{\underline{H}}}\left[\mathbf{C_r}[0]_{m,i}\right]=0$ as $h_{m,u}[l]$, $h_{i,u}[l]$, $w_m[n]$ and $w_i[n]$ are uncorrelated for $m\neq i$. The numerator of the $\gamma_u[k]'$ expression can be found as
\begin{align}
|\mathbb{E}\left[g_u[k]'\right]|^2=\biggr\rvert\mathbb{E}\left[\rho_d\sqrt{N}\mathbf{\hat{b}}_u[k]^H\mathbf{A}{\mathbf{\hat{h}}_u}[k]\right]\biggr\rvert^2&\approx\biggr\rvert\rho_d\sqrt{N}G\mathbb{E}\left[\mathbf{\hat{b}}_u[k]^H{\mathbf{\hat{h}}_u}[k]\right]\biggr\rvert^2\label{eqn:uncond_exp_1}\\
\label{eqn:uncond_exp_3}
&=|\rho_d\sqrt{N}G|^2=\rho_d^2NG^2,
\end{align}
where the approximation in (\ref{eqn:uncond_exp_1}) is due to the approximation in (\ref{eqn:rec_pow}), and (\ref{eqn:uncond_exp_3}) is owing to ZF combining. What remains is to find the denominator terms of $\gamma_u[k]'$ as follows:
\begin{align}
\text{Var}[g_u[k]']+\text{Var}[w_u[k]]=&\text{Var}[g_u[k]']+\text{Var}[{i}_u[k]]+\text{Var}[{n}_u[k]]+\text{Var}[{q}_u[k]],\label{eqn:est_total_var}
\end{align}
\begin{align}
\text{Var}[{n}_u[k]]=\mathbb{E}[|\mathbf{{\hat{b}}}_u[k]^H\mathbf{A}\tilde{\mathbf{w}}[k]|^2]&\approx G^2\mathbb{E}\left[\mathbf{{\hat{b}}}_u[k]^H\tilde{\mathbf{w}}[k]\tilde{\mathbf{w}}[k]^H\mathbf{\hat{b}}_u[k]\right]\nonumber\\
&=G^2\mathbb{E}\left[\text{Tr}\left[\tilde{\mathbf{w}}[k]\tilde{\mathbf{w}}[k]^H\mathbf{\hat{b}}_u[k]\mathbf{{\hat{b}}}_u[k]^H\right]\right]\nonumber\\
&=G^2\text{Tr}\left[\mathbb{E}\left[\mathbf{\tilde{\mathbf{w}}}[k]\tilde{\mathbf{w}}[k]^H\right]\mathbb{E}\left[\mathbf{\hat{b}}_u[k]\mathbf{\hat{b}}_u[k]^H\right]\right]\nonumber\\
&=G^2NN_o\mathbb{E}\left[\text{Tr}\left[\mathbf{\hat{b}}_u[k]\mathbf{\hat{b}}_u[k]^H\right]\right]\nonumber\\
&=G^2NN_o\mathbb{E}\left[||\mathbf{\hat{b}}_u[k]||^2\right]\approx\dfrac{G^2NN_o}{(M-U)(1-\sigma_e^2)}\label{eqn:noise_var},
\end{align}
where in the last step, the approximation $\mathbb{E}\left[||\mathbf{\hat{b}}_u[k]||^2\right]\approx 1/\left((M-U)(1-\sigma_e^2)\right)$, is used, whose proof can be found in \cite{Marzetta1088544}. $\text{Var}[{q}_u[k]]$ can be found similarly as
\begin{equation}
\label{eqn:quant_err_var_freq}
\text{Var}[{q}_u[k]|\tilde{s}_u[k]]\approx N(2-4\pi)/\left(\left(M-U\right)\left(1-\sigma_e^2\right)\right),
\end{equation}
under the approximation $\mathbf{C}_{\mathbf{q}[k]}\approx N(2-4\pi)\mathbf{I}$, which is accurate under two conditions. The first is when $L$ is large, for which the approximation $\mathbf{C_r}[0]\approx G\mathbf{I}$ has been shown to be accurate, which along with (\ref{eqn:dist_simplified}) and (\ref{eqn:arcsine_rule}) implies that $\mathbf{C_q}[0]\approx(2-4/\pi)\mathbf{I}$ is accurate. The second condition is when the oversampling rates are low and when $\rho_d^2\approx\rho_i^2$, in which case it can be shown that the approximation $\mathbf{C_r}[\ell]\approx\mathbf{0}$ for $\ell\neq0$ is accurate, which implies along with (\ref{eqn:dist_simplified}) and (\ref{eqn:arcsine_rule}) that the approximation $\mathbf{C_q}[\ell]\approx\mathbf{0}$ for $\ell\neq0$ is accurate. Then, it follows from Proposition \ref{prop:error_cov_freq} that the approximation $\mathbf{C}_{\mathbf{q}[k]}\approx N(2-4/\pi)$ is accurate. For the proof of the approximation $\mathbf{C_r}[\ell]\approx\mathbf{0}$ for $\ell\neq0$ being accurate for low oversampling rates, consider the received signal vector at the $m^{th}$ antenna, namely $\mathbf{r}_m\triangleq[r_m[0] \ r_m[1] \ \ldots r_m[N-1]]^T$. It can be written as $\mathbf{r}_m=\mathbf{\underline{H}}_m\mathbf{\underline{s}}+\mathbf{\underline{w}_m}$, where $\mathbf{\underline{H}}_m$ is a block circulant channel matrix whose $k^{th}$ row is equal to the $(mk)^{th}$ row of $\mathbf{\underline{H}}$ and $\mathbf{\underline{w}_m}$ is a vector whose $k^{th}$ element is equal to the $(mk)^{th}$ element of $\mathbf{\underline{w}}$. Moreover, defining $\mathbf{\underline{\tilde{s}}}\triangleq(\mathbf{F}_N\otimes\mathbf{I}_{U+I})\mathbf{\tilde{s}}$, where $\otimes$ represents the Kronecker product, $\mathbf{F}_N^H$ is the $N\times N$ DFT matrix, the autocorrelation of $\mathbf{r}_m$ conditioned on $\mathbf{\underline{H}_m}$, namely $\mathbf{C}_{\mathbf{r}_m}\triangleq\mathbb{E}[\mathbf{r}_m\mathbf{r}_m^H|\mathbf{\underline{H}_m}]$, can be found as
\begin{equation}
\label{eqn:r_m_autocorr}
\mathbf{C}_{\mathbf{r}_m}=\mathbf{\underline{H}}_m\mathbf{{P}}\mathbf{\underline{H}}_m^H+N_o\mathbf{I},
\end{equation}
where $\mathbf{{P}}\triangleq \mathbf{\underline{F}}^H\mathbb{E}[\mathbf{\underline{\tilde{s}}}\mathbf{\underline{\tilde{s}}}^H]\mathbf{\underline{F}}$ and $\mathbf{\underline{F}}^H=(\mathbf{F}_N^H\otimes\mathbf{I}_{U+I})$. When $\rho_d^2\approx\rho_i^2=\rho^2$, the magnitude of the element of matrix $\mathbf{\underline{P}}$ at its $m^{th}$ row and $n^{th}$ column, denoted by $|\mathbf{\underline{P}}_{m,n}|$, can be written as
\begin{align}
|\mathbf{\underline{P}}_{m,n}|=\left\lvert\frac{\rho^2}{N}\sum_{k=0}^{N-1}\mathrm{E}\left[|\tilde{s}_u[k]|^2\right]e^{-j2\pi \ell k/N}\right\rvert&\myeq
\left\lvert-\frac{\rho^2}{N}\sum_{k\notin(\mathcal{K}_D\cup\mathcal{K}_I)}\mathrm{E}\left[|\tilde{s}_u[k]|^2\right]e^{-j2\pi \ell k/N}\right\rvert\label{eqn:cor_mtx_rec_signal}\\&< \frac{\rho^2}{N}(N-|\mathcal{K}_D|-|\mathcal{K}_I|),
\end{align}
where $u$ is the user index determined by the value of $m$ and $\ell=m-n$. Moreover, the equality (*) holds when $\ell\neq 0$. Since $m^{th}$ diagonal element of $\mathbf{\underline{P}}$, namely $\mathbf{\underline{P}}_{m,m}=\rho^2/N \ \forall m$, the ratio of the magnitude of any non-diagonal element of $\mathbf{\underline{P}}$ to any diagonal element is bounded by $(N-|\mathcal{K}_D|-|\mathcal{K}_I|)$. Therefore, as $|\mathcal{K}_D|+|\mathcal{K}_I|$ approaches $N$, which occurs in a low oversampling rate scenario, the error in approximating $\mathbf{\underline{P}}$ as a diagonal matrix goes to zero. In this case, $\mathbf{C}_{\mathbf{r}_m}$, whose each element is a weighted summation of the channel coefficients $h_{m,u}[\ell]$, converges to $\mathrm{E}_{\mathbf{\underline{H}}_m}\left[\mathbf{C}_{\mathbf{r}_m}\right]=\mathrm{E}_{\mathbf{\underline{H}}_m}\left[\mathbf{\underline{H}}_m\mathbf{\underline{H}}_m^H\right]\mathbf{\underline{P}}+N_o\mathbf{I}=(U+I)\mathbf{\underline{P}}+N_o\mathbf{I}$ as $L$ grows large, which can be shown rigorously following similar steps as in (\ref{eqn:mean_cond_H})-(\ref{eqn:chebyshev_ineq}). Since the proof is the same $\forall m$, the error in approximating $\mathbf{C}_{\mathbf{r}_m}$ matrices as a diagonal matrices $\forall m$, equivalently approximating $\mathbf{C_r}[\ell]\approx\mathbf{0}$ for $\ell\neq0$ or $\mathbf{C}_{\mathbf{q}[k]}\approx N(2-4/\pi)\mathbf{I}$, goes to zero as $L$ grows large and as $|\mathcal{K}_D|+|\mathcal{K}_{I}|$ approaches $N$ (which is a case for low oversampling rates) when $\rho_d^2\approx\rho_i^2$.

The proof continues with obtaining $\text{Var}\left[i_u[k]\right]$ as follows:
\begin{align}
\text{Var}\left[i_u[k]\right]=&\rho_d^2N\text{Var}\left[\sum_{z\neq u, z\in\mathcal{U}_d}\mathbf{{\hat{b}}}_u[k]^H\mathbf{A}{\mathbf{\hat{h}}_z}[k]\tilde{s}_z[k]+\sum_{z\neq u, z\in\mathcal{U}_d}{\mathbf{\hat{b}}_u}[k]^H\mathbf{A}\mathbf{\tilde{e}}_z[k]\tilde{s}_z[k]\right]\nonumber\\
\approx& G^2\rho_d^2N\sum_{z\neq u, z\in\mathcal{U}_d}\mathbb{E}\left[|\mathbf{{\hat{b}}}_u[k]^H{\mathbf{\tilde{h}}_z}[k]|^2\right]+\sum_{z\neq u, z\in\mathcal{U}_d} G^2\rho_d^2N\mathbb{E}\left[|\mathbf{\tilde{e}}_z[k]^H{\mathbf{\hat{b}}_u}[k]|^2\right].\label{eqn:rec_sig_compnts_append_2}
\end{align}
Here, $\mathbb{E}\left[|\mathbf{{\hat{b}}}_u[k]^H{\mathbf{\tilde{h}}_z}[k]|^2\right]=0$ $\forall z\neq u$ due to ZF combining and
%\begin{align}
%\mathbb{E}[&|\mathbf{{\hat{b}}}_u[k]^H{\mathbf{\tilde{h}}_z}[k]|^2]\\
%=&\mathbb{E}\left[\mathbf{{\hat{b}}}_u[k]^H{(\mathbf{\hat{h}}_z}[k]+\mathbf{\tilde{e}}_z[k])({\mathbf{\hat{h}}_z}[k]^H+\mathbf{\tilde{e}}_z[k]^H)\mathbf{{\hat{b}}}_u[k]\right]\\
%=&\mathbb{E}\left[\mathbf{{\hat{b}}}_u[k]^H\mathbf{\tilde{e}}_z[k]\mathbf{\tilde{e}}_z[k]^H\mathbf{{\hat{b}}}_u[k]\right]\\
%=&\mathbb{E}\left[\text{Tr}\left[\mathbf{{\hat{b}}}_u[k]^H\mathbf{\tilde{e}}_z[k]\mathbf{\tilde{e}}_z[k]^H\mathbf{{\hat{b}}}_u[k]\right]\right]\\
%=&\mathbb{E}\left[\text{Tr}\left[\mathbf{\tilde{e}}_z[k]\mathbf{\tilde{e}}_z[k]^H\mathbf{{\hat{b}}}_u[k]\mathbf{{\hat{b}}}_u[k]^H\right]\right]\\
%=&\text{Tr}\left[\mathbb{E}\left[\mathbf{\tilde{e}}_z[k]\mathbf{\tilde{e}}_z[k]^H\right]\mathbb{E}\left[\mathbf{{\hat{b}}}_u[k]\mathbf{{\hat{b}}}_u[k]^H\right]\right]\\
%=&\sigma_e^2||\mathbf{{\hat{b}}}_u[k]||^2\\
%\approx &\dfrac{\sigma_e^2}{(M-K)(1-\sigma_e^2)}\label{eqn:var_term_inside_sum}.
%\end{align}
\begin{align}
\mathbb{E}\left[|\mathbf{\tilde{e}}_z[k]^H\mathbf{\hat{b}}_u[k]|^2\right]&=\mathbb{E}\left[\text{Tr}\left[\mathbf{\hat{b}}_u[k]\mathbf{\hat{b}}_u[k]^H\mathbf{\tilde{e}}_z[k]\mathbf{\tilde{e}}_z[k]^H\right]\right]\nonumber\\
&=\text{Tr}\left[\mathbb{E}\left[\mathbf{\hat{b}}_u[k]\mathbf{\hat{b}}_u[k]^H\right]\mathbb{E}\left[\mathbf{\tilde{e}}_z[k]\mathbf{\tilde{e}}_z[k]^H\right]\right]\label{eqn:indep_chan_est_error}\\
&=\sigma_e^2\mathbb{E}\left[||\mathbf{{\hat{b}}}_u[k]||^2\right]\approx \dfrac{\sigma_e^2}{(M-U)(1-\sigma_e^2)}\label{eqn:chan_est_error_var_term},
\end{align}
where (\ref{eqn:indep_chan_est_error}) is due to the uncorrelatedness of the channel estimation error $\mathbf{\tilde{e}}_u[k]$ with the channel estimate $\mathbf{\hat{h}}_z[k]$, hence with $\mathbf{\hat{b}}_z[k]$. Using (\ref{eqn:rec_sig_compnts_append_2}) and (\ref{eqn:chan_est_error_var_term}), it can be written that
\begin{align}
\text{Var}\left[i_u[k]\right]=\text{Var}\left[g_u[k]'+i_u[k]\right]\approx\dfrac{G^2\rho_d^2N(U-1)\sigma_e^2}{(M-U)(1-\sigma_e^2)}\label{eqn:interf_variance}.
\end{align}
Similarly, $\text{Var}\left[g_u[k]'\right]$ can be found as 
\begin{equation}
\label{eqn:gain_var}
\text{Var}\left[g_u[k]'\right]\approx {\sigma_e^2}{(M-U)/(1-\sigma_e^2)}.
\end{equation}
Then, from (\ref{eqn:est_total_var}), (\ref{eqn:noise_var}), (\ref{eqn:quant_err_var_freq}), (\ref{eqn:interf_variance}) and (\ref{eqn:gain_var}) it follows that
\begin{align}
\text{Var}[g_u[k]']+&\text{Var}[\hat{i}_u[k]+\hat{n}_u[k]+\hat{q}_u[k]]\approx\dfrac{G^2NN_o+N(2-4/\pi)+G^2\rho_d^2NU\sigma_e^2}{(M-U)(1-\sigma_e^2)},
\end{align}
which implies the proposition statement along with (\ref{eqn:uncond_exp_3}).\hfill \qedsymbol
\bibliographystyle{IEEEtran}
\bibliography{IEEEabrv,referans}

\end{document}